\documentclass[a4paper,11pt]{article}
\usepackage{jcappub} 
\usepackage{lineno}
\usepackage{macros}

\usepackage{ulem}
\title{\boldmath Magnetogenesis in Low-Scale Inflation: Viability for Astrophysical Magnetic Fields}
\author[a,1]{Ramkishor Sharma,}
\author[a,2]{Dilip Kumar,}
\author[b,c,3]{Kandaswamy Subramanian,}
\author[a,4]{and Soma Sanyal}
\affiliation[a]{School of Physics, University of Hyderabad, Central University P.O, Hyderabad-500046, Telangana, India.}
\affiliation[b]{IUCAA, Post Bag 4, Ganeshkhind, Pune 411007, India}
\affiliation[c]{Department of Physics, Ashoka University, Rajiv Gandhi Education City, Rai, Sonipat 131029, Haryana, India}

\emailAdd{ramkishor.sharma@uohyd.ac.in$^{1}$}
\emailAdd{21phph03@uohyd.ac.in$^{2}$}
\emailAdd{kandu@iucaa.in$^{3}$}
\emailAdd{somasanyal@uohyd.ac.in$^{4}$}

\abstract{The indirect evidence for magnetic fields in the intergalactic medium motivates the study of their possible primordial origin, including generation during inflation. One mechanism for inflationary magnetogenesis involves a pseudoscalar inflaton coupled to a gauge field through the interaction $g_{\phi\gamma}\phi F_{\mu\nu}\tilde{F}^{\mu\nu}$. 
In this work, we study the viability of this mechanism in the context of low scale inflation, 
particularly on inflationary energy scales near the electroweak (EW) scale. We find that 
the post inflationary amplification of the gauge fields can produce astrophysically relevant magnetic fields for $g_{\phi\gamma}\ge 360~m_{\rm pl}^{-1}$ in the absence of Schwinger backreaction for an inflationary scale around the EW scale. We then include the backreaction of the charged current generated through the Schwinger mechanism. We find that the induced current suppresses the gauge field amplification and leads to a saturation of the gauge field energy density at a value of order $m_e^4$, where $m_e$ is the electron mass. Assuming this saturation level, we find that magnetic fields above the lower bound inferred from the non detection of secondary GeV photons from blazars can be obtained when reheating occurs at temperatures below $\sim 0.3$ GeV. For an inflationary energy scale of $100~{\rm GeV}$, achieving this saturation until a reheating temperature, $T\le 0.3~{\rm GeV}$, requires a coupling strength $g_{\phi\gamma}\ge 20000~m_{\rm pl}^{-1}$. We obtain a present-day magnetic field strength of $B_0 \sim 2\times10^{-13}$ G with a coherence length of $L_{0}\sim 5\times10^{-3}$ kpc for the reheating temperature of $100$ MeV.}

\begin{document}
\maketitle
\flushbottom

\section{Introduction}
\label{sec:intro}
Magnetic fields have been observed over a wide range of length scales in the Universe \cite{Widrow:2002ud,Bernet:2008qp,Beck:2008ty,Kronberg:2007dy}. The inference of their presence in cosmic void regions  \cite{Neronov:1900zz, Tavecchio:2010mk, Dolag:2010ni, Essey:2010nd, Taylor:2011bn, Takahashi:2013lba, Finke:2013tyq, Finke:2015ona,MAGIC:2022piy} has motivated studies on their possible generation in the early Universe. Several ideas have been proposed for primordial magnetogenesis, which can broadly be categorised into two classes: generation during inflation \cite{Turner:1987bw, Ratra:1991bn, Garretson:1992vt,Martin:2007ue,Demozzi:2009fu,Kanno:2009ei,Bamba:2006ga,Anber:2006xt,Campanelli:2013mea,Ferreira:2013sqa,Ferreira:2014hma,Ng:2015ewp,Fujita:2015iga,Adshead:2016iae,Kushwaha:2020nfa,Tripathy:2022iev} and generation during phase transitions \cite{Vachaspati:1991nm, Sigl:1996dm, Dolgov:2001nv, Stevens:2007ep, DiazGil:2007dy,Kahniashvili:2009qi, Henley:2010ba,Grasso:2000wj,Cheng:1994yr} (see Refs.~\cite{Durrer:2013pga,Subramanian:2015lua,Vachaspati:2020blt} for reviews).  

The motivation for inflationary magnetogenesis arises from the fact that if magnetic fields are generated during inflation, they naturally extend over superhorizon scales, much like the density perturbations that later seeded the formation of large-scale structures in the Universe. However, within standard electromagnetic (EM) theory, magnetic field generation during inflation is not possible due to conformal invariance. To overcome this, conformal symmetry must be broken, for example, through couplings of the form $f^2 (\phi) F_{\mu\nu}F^{\mu\nu}$ \cite{Turner:1987bw,Ratra:1991bn,Demozzi:2009fu,Martin:2007ue,Ferreira:2013sqa,Nandi:2017ajk,Sharma:2017eps,Kushwaha:2020nfa,Tripathy:2022iev}, where $\phi$ denotes a scalar field present during inflation, which could be the inflaton itself or any other scalar field present during the inflationary epoch. These models, however, suffer from the strong coupling and backreaction problems. A viable alternative was proposed in Refs.~\cite{Sharma:2017eps,Sharma:2018kgs}, where the generation of magnetic fields of astrophysical interest requires reheating to occur at scales below $10^4 \, {\rm GeV}$.  

Another possibility arises if inflation is driven by a pseudoscalar\footnote{In this article, we use the terms pseudoscalar and axion interchangeably.} field coupled to the $U(1)$-gauge sector via a term of the form\footnote{Recently, the generation of magnetic fields in the post-recombination era through a similar mechanism has been explored, where the pseudoscalar field acts as an ultralight dark matter candidate \cite{Brandenberger:2025gks, Brahma:2026fre,Kushwaha:2026aab,Sharma:2026hzp,Gerlach:2026sga}.} $\phi F_{\mu\nu}\tilde{F}^{\mu\nu}$\cite{Garretson:1992vt,Anber:2006xt,Fujita:2015iga,Adshead:2015pva,Adshead:2016iae,Kamarpour:2018ckk,Sobol:2019xls,Gorbar:2021zlr,Gorbar:2021rlt,Brandenburg:2024awd,Sharma:2024nfu}. As the $\phi$ field rolls down its potential, one of the two helicity modes of the gauge field experiences a tachyonic instability, leading to its amplification. The amount of amplification depends on the velocity of the $\phi$ field and the strength of the coupling. Recent studies suggest that it is difficult to generate magnetic fields of astrophysical interest in such models if inflation occurs at high scales \cite{Iarygina:2025ncl,Gorbar:2021rlt} due to the suppression in the obtained strength via the charged current produced by the Schwinger mechanism. 
An additional problem for studying high-scale magnetogenesis comes from the fact that if magnetic fields are generated before the electroweak (EW) phase transition, those strong enough to explain the lower bound from blazars, could
overproduce baryon isocurvature perturbations, which are constrained by Big Bang nucleosynthesis \cite{Kamada:2016cnb,Kamada:2020bmb}. However, recent studies have revisited the baryon-isocurvature constraints and found that these constraints do not completely exclude the possibility of generating primordial magnetic fields before the EW era \cite{Boyer:2025jno,Hamada:2025ooy}. Independently of this issue, the effects of the charged current induced by Schwinger pair production on gauge field amplification need to be taken into account when assessing the viability of these scenarios. A possible magnetogenesis scenario in low-scale inflation that avoids the baryon isocurvature problem was studied in Ref.~\cite{Yanagihara:2023qvx}. However, this study does not take into account the effects of the charged current due to the Schwinger mechanism.

In this paper, we investigate whether magnetic fields of astrophysical interest can be generated in a low-scale inflation model driven by a pseudoscalar field coupled to the $U(1)$ gauge sector via the term $\phi F_{\mu\nu}\tilde{F}^{\mu\nu}$, considering the Generalised Minimal Supersymmetric Standard Model (GMSSM) as a representative low-scale inflation scenario. In low scale inflationary scenarios, gauge-field production during inflation is negligible because of the small velocity of the inflaton field. Instead, most of the gauge field amplification occurs during the post inflationary phase, when the inflaton undergoes coherent oscillations around the minimum of its potential. Further, we incorporate the electric current generated by charged particles through the Schwinger mechanism and study its impact on the gauge field evolution. The paper is organised as follows. In \Sec{section2}, we review the basic mechanism of magnetic-field generation through the $\phi F\tilde{F}$ interaction and study the post-inflationary amplification of the gauge field during the oscillatory behaviour of the $\phi$ field in GMSSM inflation. In \Sec{sec_cond}, we investigate the evolution of the gauge field after incorporating the charged current produced via the Schwinger mechanism. In \Sec{section4}, we examine the possibility of generating magnetic fields of astrophysical interest in the GMSSM and compare the resulting field strengths with the lower bounds inferred from the non-detection of secondary GeV photons from blazars. Finally, in \Sec{section5}, we summarise our results.

\section{Gauge field dynamics in a low-scale inflation model}\label{section2}
The dynamics of the pseudoscalar coupled with a U(1) gauge field is described by the action;
\begin{align}\label{Action_axion_U1}
    S=\int d^4x \sqrt{-g} \Bigg[\frac{m_\mathrm{pl}^2}{16 \pi}R-\frac{1}{2}\partial_\mu\phi \partial^\mu\phi-V(\phi)-\frac{1}{4}F_{\mu \nu}F^{\mu\nu}-\frac{g_{\phi\gamma}}{4}\phi F_{\mu \nu}\tilde{F}^{\mu\nu}\Bigg].
\end{align}
Here, $\phi$ and $V(\phi)$ represent the pseudoscalar field and its potential, respectively, $F_{\mu\nu}\equiv \partial_{\mu} A_{\nu}-\partial_{\nu} A_{\mu}$ denotes the gauge field strength tensor and $\tilde{F}^{\mu\nu}$ is its dual tensor, $g_{\phi \gamma}$ is the coupling strength between pseudoscalar and gauge field. Assuming that the background evolution is described by a spatially flat Friedmann-Lemaitre-Robertson-Walker (FLRW) metric\footnote{We use the $(-,+,+,+)$ metric convention.}, we get the following equation of motion for the pseudoscalar and the gauge field;
\begin{align}
    \Ddot{\phi}+3 H \dot{\phi}-\frac{1}{a^2}\ddel \phi+ \frac{d V}{d \phi}&=g_{\phi\gamma}\,\frac{1}{a^4}\,\EE\cdot\BB\label{phi_eqn}\,,\\
    \ddot{\AA}+H \dot{\AA}-\frac{1}{a}\Del \dot{A_0}-\frac{1}{a^2}\ddel \AA+\frac{1}{a^2}\Del(\Del \cdot \AA)&
    -g_{\phi \gamma}\left(\frac{\dot{\phi}\BB}{a}+\frac{1}{a^2}\Del \phi \times \EE\right)=0.\,\label{Ai_eqn}
\end{align}
Here, dot ($\dot{}$) on the top of the variable represents the derivative with respect to cosmic time ($t$), $\EE=-a \dot{\AA}+\del A_0$ and $\BB=\del \times \AA$ denote the electric and magnetic parts of the gauge field, and $a(t)$ is the scale factor. We define $\nabla^2\equiv\delta^{ij}\partial_i\partial_j$, where $\delta^{ij}$ is the Kronecker delta. The initial condition of the $\phi$ is chosen such that its spatial fluctuations are negligible compared to the homogeneous part. At a later time during its evolution, these spatial fluctuations can be amplified by the backreaction of the gauge fields. By assuming that we are working in a coupling regime where the backreaction of the gauge field on the pseudoscalar dynamics can be neglected, we get the following evolution equations\cite{Barnaby:2011vw,Fujita:2015iga};
\begin{align}
    \Ddot{\phi} + 3 H \Dot{\phi} + \frac{\partial V}{\partial \phi} = 0, \quad \,& H^{2} = \frac{8 \pi}{3 m_{\rm pl}^2} \bigg( \frac{1}{2} \Dot{\phi}^{2} + V(\phi) \bigg) , \label{inflaton}\\
    \Ddot{\mathcal{A}}_{\pm} + H \dot{\mathcal{A}}_{\pm} + \frac{k}{a} &\bigg( \frac{k}{a} \mp g_{\phi\gamma}\dot{\phi} \bigg) \mathcal{A}_{\pm} = 0 . \label{A_equation}
\end{align}
Here, $H\equiv\dot{a}/a$ denotes the Hubble parameter, $m_{\rm pl}=1.22\times10^{19}~{\rm GeV}$ is the Planck mass, $k$ is the wavenumber and $\mathcal{A}_{\pm}$ represents the Fourier modes of $\AA$ in circular polarisation basis. Assuming negligible axion inhomogeneities, $A_0$ vanishes in the Coulomb gauge.
As a representative low-scale inflation model, we consider the GMSSM model, for which the inflaton potential is given by \cite{Allahverdi:2006iq,Allahverdi:2007wb,Martin:2013nzq},
\begin{align}
V(\phi)=\Lambda^4 \Bigg(\frac{1}{2}\left(\frac{\phi}{\phi_0}\right)^2-\frac{\alpha_G}{3}\left(\frac{\phi}{\phi_0}\right)^6+\frac{\alpha_G}{10}\left(\frac{\phi}{\phi_0}\right)^{10}\Bigg).
\end{align}
Here, $\Lambda$ represent the energy scale of inflation and $\alpha_G$ is a constant which decides the shape of $V(\phi)$ around the point $\phi_0$. We note that obtaining a sufficiently long period of inflation requires significant fine-tuning of the parameter $\alpha_G$. In particular, requiring an inflationary duration $\Delta N$ leads to the condition (see equation (6.116) in Ref.~\cite{Martin:2013nzq}),
\begin{align}
\alpha_G > 1-\left(\frac{\phi_0}{\sqrt{8\pi}m_{\rm pl}}\right)^4\frac{\pi^2}{900 \Delta N^2}.
\end{align}
Here, $\Delta N$ denotes the duration of inflation. For $\Delta N=60$ and $\phi_0=10^{-5}m_{\rm pl}$, this condition implies $\alpha_G > 1-10^{-23}$. Working with such an extremely fine-tuned value of $\alpha_G$ is numerically challenging. However, our primary interest is in the amplification of the gauge fields, which takes place predominantly post-inflation in a low scale inflation model as discussed below, rather than in the dynamics during inflation. The inflaton dynamics post inflation are expected to be only weakly sensitive to the value of $\alpha_G$, as $\phi$ becomes smaller than $\phi_0$, and the terms in the potential containing $\alpha_G$ become subdominant compared to those without $\alpha_G$. Consequently, we expect the gauge-field amplification to be largely insensitive to the particular value of $\alpha_G$ adopted in our numerical calculations. We use $1-\alpha_G\sim10^{-8}$ in our simulations, which allows us to avoid the numerical difficulties associated with the extremely fine tuned value required during inflation.

As the $\phi$ field rolls down its potential, it leads to the amplification of the gauge field. This phenomenon has been studied in detail both analytically \cite{Garretson:1992vt,Barnaby:2010vf,Anber:2009ua,Sobol:2019xls,Adshead:2016iae,Gorbar:2021rlt} and numerically \cite{Caravano:2022epk,Figueroa:2023oxc,Figueroa:2024rkr,Sharma:2024nfu}. As this amplification is dependent on the $\phi$ velocity, the maximum amplification happens near the end of inflation. In this paper, we study the dynamics of the inflaton field $\phi$ and the gauge field in a low-scale inflationary scenario, with an inflationary energy scale of around $100$ GeV. In low-scale inflation, the inflaton rolls with a much smaller velocity compared to high-scale models in order to match the observed amplitude of curvature perturbations. Because $\dot{\phi}$ is significantly smaller, the gauge field does not experience any substantial amplification during inflation for the same values of $g_{\phi\gamma}$ that lead to amplification in the high scale quadratic inflation model \cite{Adshead:2016iae,Sharma:2024nfu}. This is because the instability term, $k g_{\phi\gamma} \dot{\phi}/a$, remains too small to trigger tachyonic growth. Consequently, in this scenario, we expect the gauge field to be amplified only in the post-inflationary era. From \Eq{A_equation}, it follows that modes satisfying $k^2/a^2<k g_{\phi\gamma} \dot{\phi}/a$ experience a tachyonic instability during a fraction of each oscillation cycle of $\dot{\phi}$ post inflation. As a result, the two helicity modes, $\mathcal{A}_{+}$ and $\mathcal{A}_{-}$, are amplified during alternate half-cycles of the $\phi$ oscillation. The range of $k$-modes for which the tachyonic instability occurs is given by,
\begin{equation}\label{kc_def}
    k \leq g_{\phi\gamma} a H m_{pl} \sqrt{\frac{3}{4\pi}} \equiv k_c \qquad \text{or} \qquad \frac{k}{a H}\leq 0.5 g_{\phi \gamma} m_{\rm pl}.
\end{equation}
To obtain the above expression, we assume that, during the post-inflationary era, $\dot{\phi}=\sqrt{3/4\pi} H m_{\rm pl}$. In the post-inflationary era, $\phi$ undergoes coherent oscillations, and $\dot{\phi}=\sqrt{3/4\pi} H m_{\rm pl}$ is the amplitude of these oscillations. This can be obtained from the first Friedmann equation, $H^2=8\pi/(3m_{\rm pl}^2)(\dot{\phi}^{2}/2+V(\phi))$ by considering the point during the oscillation when $V(\phi)\simeq 0$ and the energy density is dominated by the kinetic term. The wavenumber $k_c \simeq 0.5 g_{\phi\gamma} m_{\rm pl}aH$ undergoes the maximum amplification, and the resulting gauge field energy density is therefore expected to peak around this mode. Accordingly, in the numerical simulations presented in the next section, we choose a range of wavenumbers around $k_c|_e \simeq 0.5~ g_{\phi\gamma} m_{\rm pl}a_e~H_e$ to ensure that the peak of the generated gauge field energy density spectrum is well resolved. Here, $a_e$ and $H_e$ represent the value of the scale factor and the Hubble parameter at the end of inflation.

\begin{figure}
    \centering
    \includegraphics[width=1\linewidth]{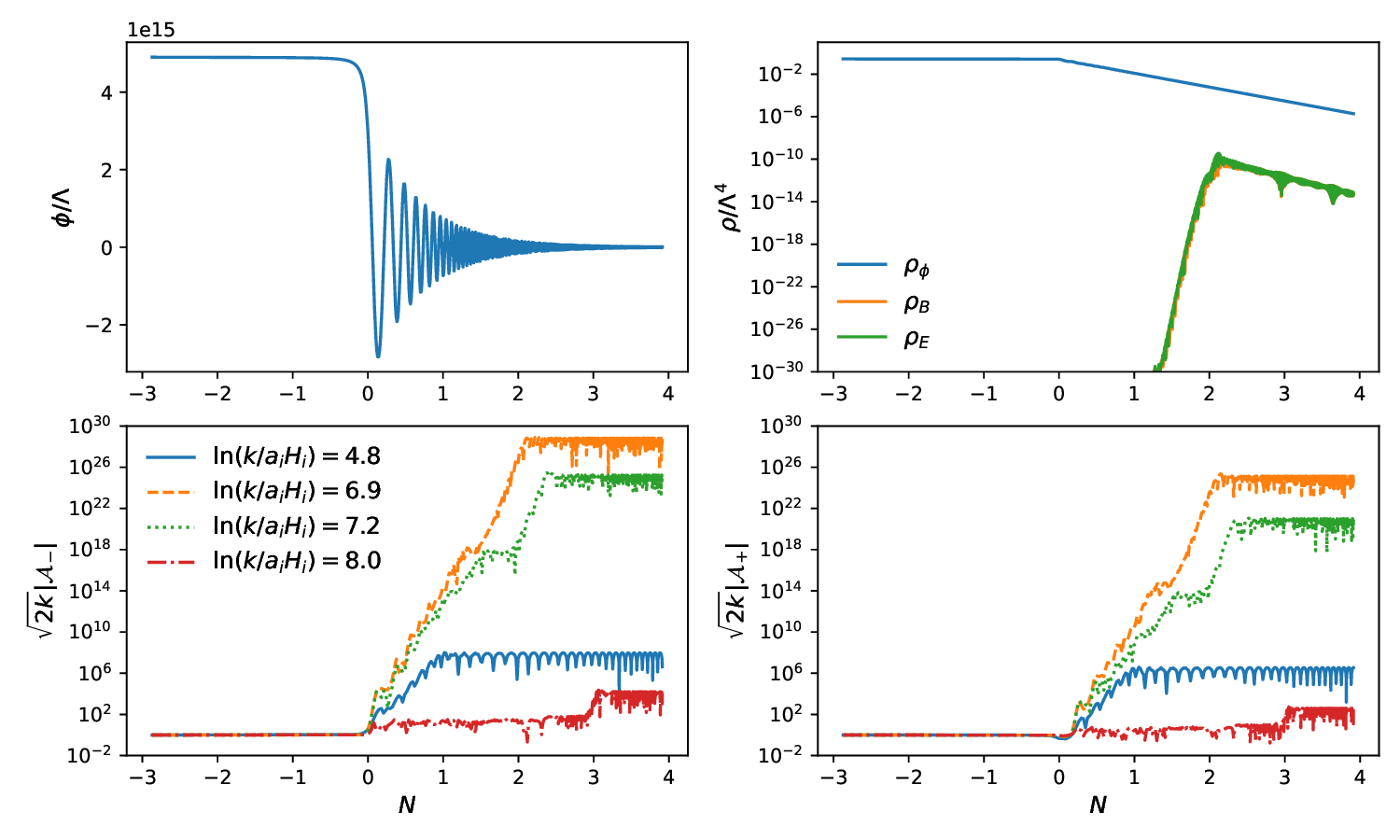}
    \caption{\textbf{Upper panels:} Evolution of the inflaton field $\phi$ (left panel) and the energy densities $\rho_{\phi}$ in blue curve, $\rho_E$ in green cruve, and $\rho_B$ in yellow curve (right panel) as functions of the number of e-fold, $N$, for the quadratic inflation model (run A). \textbf{Lower panel:} Evolution of the vector potential modes, $\sqrt{2k}|A_-|$ (left) and $\sqrt{2k}|A_+|$ (right) for $\ln (k/(a_iH_i))=4.8,~6.9,~7.2$, and $8.0$. Here, the yellow curve corresponds to the mode $k=k_c$ defined in \Eq{kc_def}. This figure is for run A1, and the values of the parameters for this run are given in \Tab{table1}. }
    \label{gmssm_plot}
\end{figure}

To study the dynamics of the inflaton and the gauge field numerically, we solve  \Eqss{inflaton}{A_equation} using the \texttt{axionU1back.f90} module of the {\sc Pencil Code}~\cite{JOSS}. We initiate our simulation $N_e$ e-folds before the end of inflation. For the gauge fields evolution, we sample 256 $k$-modes in the range $e^{n_{\min}} aH \le k \le e^{n_{\max}} aH$. The values of the parameter taken in our numerical runs are summarised in \Tab{table1}. The values of $n_{min}$ and $n_{max}$ have been chosen such that the peak of the gauge field, approximately at the wavenumber $\sim 0.5 g_{\phi\gamma} m_{\rm pl}a_e~H_e$ spectrum is well resolved. At the beginning of our simulation, we assume the gauge field are in the Bunch-Davies vacuum and the initial value of $\mathcal{A}_\pm$ and $\dot{\mathcal{A}}_\pm$ for a particular $k-$mode is given by,
\begin{align}
    \mathcal{A_\pm}=\frac{a_i}{\sqrt{2k}}e^{\frac{ik}{a_i H_i}}, \quad \quad
    \dot{\mathcal{A}}_{\pm}=\frac{i k}{\sqrt{2k}}e^{\frac{ik}{a_i H_i}}.
\end{align}
Here, $a_i$ and $H_i$ denote the value of the scale factor and the Hubble parameter at the beginning of our simulation.
\begin{table}
\begin{center}
\begin{tabular}{p{.6cm}|p{1.25cm}|p{1.2
cm}|p{1.8cm}|p{0.6cm}|p{.6cm}|p{.7cm}|p{.8cm}|p{1.6cm}}
\hline
Run & $g_{\phi \gamma}$ (in~${m_\mathrm{pl}^{-1}}$) &$M/m_{\rm pl}$ & $\frac{\Lambda^2}{\phi_0 m_{\rm pl}}$ & grid size& $N_{e}$  &$n_{min}$&$n_{max}$&  \text{Model} \\
\hline
A1 & $400$ & $100$ & $1.7 \times 10^{-33}$        & $256$ & $2.87$  &$4$ &$8$& GMSSM\\ 
A2 & $365$ & $100$ & $1.7 \times 10^{-33}$        & $256$ & $2.87$  &$4$ &$8$& GMSSM\\ 
A3 & $420$ & $100$ & $1.7 \times 10^{-33}$        & $256$ & $2.87$  &$4$ &$8$& GMSSM\\ 

\hline
\end{tabular}
\caption{The model parameter for different scenarios of GMSSM inflation.}\label{table1}
\end{center}
\end{table}
We show the evolution of $ \phi$ (upper left panel), energy densities (upper right panel) and the gauge-field modes (bottom panels) in \Fig{gmssm_plot} for the run A1. The energy densities are defined as,
\begin{align}\label{energy_densities}
    \rho_{\phi}=\frac{1}{2}\dot{\phi}^2+V(\phi), \quad \quad
    &\rho_B \equiv\int d\ln k \frac{d \rho_B}{d \ln k}=\frac{1}{a^4} \int d \ln k \frac{k^5}{(2\pi)^2 }(|\mathcal{A}_+|^2+|\mathcal{A}_-|^2),\\
    \rho_E &\equiv \int d\ln k \frac{d \rho_E}{d \ln k}=\frac{1}{a^4} \int d \ln k \frac{k^3}{(2\pi)^2 }(|\dot{\mathcal{A}}_+|^2+|\dot{\mathcal{A}}_-|^2).
\end{align}
In this figure, $N=0$ denotes the end of inflation, which is marked by the first slow-roll parameter, $\epsilon\equiv -\dot{H}/H^2=1$. As shown in the upper left panel, the $\phi$ field remains nearly constant during inflation and begins to oscillate after inflation with a frequency determined by its effective mass, $m_{\phi}\equiv\Lambda^2/\phi_0$. This behaviour can be understood as follows. After inflation, the value of $\phi$ becomes much smaller than $\phi_0$. Consequently, in the potential $V(\phi)$, the terms proportional to $(\phi/\phi_0)^4$ and $(\phi/\phi_0)^6$ can be neglected compared to the quadratic term. The potential therefore reduces to $V(\phi)\simeq \frac{1}{2}\Lambda^2(\phi/\phi_0)^2$. Comparing this expression with the standard quadratic potential, $V(\phi)=\frac{1}{2}m_\phi^2\phi^2$, we identify the effective mass as $m_\phi=\Lambda/\phi_0$. The values of the effective mass for our different simulation runs are given in the fourth column of \Tab{table1}. The blue, green and yellow curves in the upper right panel show the time evolution of the inflaton energy density, electric and magnetic component of the gauge field energy density, respectively. In the lower panel of \Fig{gmssm_plot}, we show the evolution of $\sqrt{2k}\,|A_{-}|$ and $\sqrt{2k}\,|A_{+}|$ for different wavenumbers corresponding to $\ln(k/(a_iH_i))=4.8$ (blue curve), $6.9$ (yellow curve), $7.2$ (green curve), and $8.0$ (red curve). As evident, both helicity modes undergo amplification only during the post-inflationary era through successive cycles of tachyonic resonance driven by the oscillatory behaviour of the inflaton field. We also note that the amplification of the negative helicity mode begins slightly earlier than that of the positive helicity mode, leading to an overall dominance of the negative helicity mode.

\begin{figure}
    \centering
    \includegraphics[width=1\linewidth]{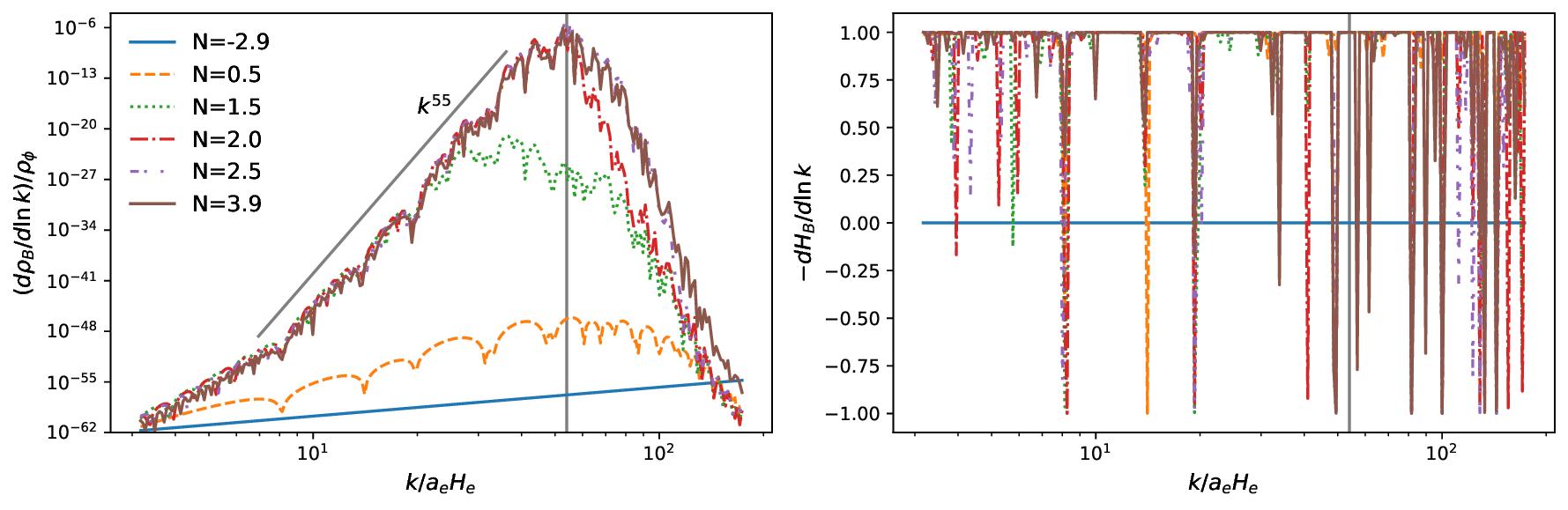}
    \caption{The magnetic energy spectra normalised by $\rho_{\phi}$ (left panel) and helicity spectra (right panel) defined in \Eqs{energy_densities}{hel_spectra} at $N=-2.9$ (solid blue curve), $N=0.5$ (dashed orange curve), $N=1.5$ (dotted green curve), $N=2.0$ (dash-dotted red curve), $N=2.5$ (dotted purple curve) and $N=3.9$ (solid brown curve) are shown for A1 run. Here, the wavenumbers on the x-axes are normalised by the comoving Hubble parameter at the end of inflation, $a_e H_e$. The vertical grey line in each panel represents a value of $k/(a_e H_e)=54$ to mark the peak of the magnetic energy spectra.}
    \label{spec_gmssm}
\end{figure}

Furthermore, we also calculate the spectra of the magnetic component of the gauge field given in \Eq{energy_densities} and it is shown in the left panel of \Fig{spec_gmssm} for $N=-2.9$, $0.5$, $1.5$, $2.0$, $2.5$, $3.9$. The spectra values in \Fig{spec_gmssm} are normalised by $\rho_{\phi}$ and the wave numbers are normalised by the commoving Hubble parameter at the end of inflation, $a_e H_e$.  The right panel of \Fig{spec_gmssm} shows the time evolution of the spectra of the fractional helicity defined as,
\begin{align}\label{hel_spectra}
    \frac{d H_B}{d \ln k}= \frac{|\mathcal{A}_+|^2-|\mathcal{A}_-|^2}{|\mathcal{A}_+|^2+|\mathcal{A}_-|^2}.
\end{align}
The vertical grey line in this plot denotes the wavenumber for which $k/a_eH_e=54$ to mark the wavenumber corresponding to the peak of the spectrum. Although the magnetic spectrum below the peak cannot be described by a single power law, fitting a power law over a limited range of wavenumbers immediately below the peak yields a very steep spectral index. The grey line corresponding to a $k^{55}$ power law in \Fig{spec_gmssm} is included merely to illustrate the steepness of the spectrum in this region. The magnetic field spectrum obtained in the present case is significantly steeper below the peak than that found in high-scale inflationary scenarios. Such a steep spectrum is expected to evolve into a much shallower one during its subsequent nonlinear evolution in the radiation-dominated era. However, this evolution occurs only on the wave numbers for which the eddy turnover time is shorter than the Hubble time.

\section{Post inflationary gauge field dynamics including the Schwinger current}\label{sec_cond}
The gauge fields produced during the post-inflationary amplification discussed in the previous section can induce the production of charged particles from the vacuum through the Schwinger mechanism \cite{2006physics...5038H,Schwinger:1951nm,Frob:2014zka,Sharma:2017ivh}. This process becomes important when the EM field strength becomes large compared to the squared mass of the charged particles. The produced charged particles generate a field-dependent electric current, which modifies the evolution of the gauge field. Once the backreaction of this induced current becomes significant, it can substantially suppress further amplification of the gauge field. The effects of the Schwinger-induced current and its backreaction on gauge-field evolution have been extensively studied in the literature \cite{Kobayashi:2014zza,Bavarsad:2016cxh,Domcke:2018eki,vonEckardstein:2024tix}.

In this section, we investigate the post-inflationary evolution of the gauge field by incorporating the effects of the charged current generated by the particles produced through the Schwinger mechanism. For this purpose, we take the following expression for the induced current, as given in equation (1.2) of Ref.~\cite{vonEckardstein:2024tix}:
\begin{align}\label{cond_exp}
   \JJ=\frac{e^3}{6 \pi^2 H} \EE |B| \coth\left(\frac{\pi |B|}{E}\right) e^{-\frac{\pi m^2}{eE}}
\end{align}
Here, $|B|$ represents the magnetic field component in the direction of $\EE$, $e$ and $m$ are the charge and the mass of the charged particle. The above expression for the charged current is derived for a homogeneous EM field in a de Sitter background. In our case, most of the gauge field production occurs during the post inflationary era, when the universe is effectively matter dominated. However, the characteristic rate of gauge field production is much larger than the Hubble rate, and the gauge field spectrum peaks at physical wavenumbers well inside the Hubble horizon, $\frac{k_c}{a_eH_e}\sim 54>1$ for $g_{\phi\gamma m_{\rm pl}}=400$. Thus, the expansion rate of the universe can be treated as slowly varying on the relevant timescales, allowing the Schwinger pair-production process to be treated locally. We therefore use the de Sitter result as a local approximation for the Schwinger induced current during the post inflationary era. We note that the validity of this approximation also requires the EM fields to vary sufficiently slowly on the characteristic scales related to pair production.

The expression in \Eq{cond_exp} for the current is valid for the case in which the electric and magnetic fields are aligned with each other. The more general case has been discussed in Ref.~\cite{vonEckardstein:2024tix}. However, in order to keep the present analysis simple, we restrict ourselves to this configuration. Moreover, for almost fully helical fields, as we find here, the aligned case is a good approximation. In our analysis, we assume that the dominant contribution to the Schwinger current arises from the production of the lightest charged fermions in the Standard Model, namely electrons and positrons. We therefore take $m = 0.511~{\rm MeV}$ in our analysis.

\subsection{Numerical implementation}
After including the Schwinger-induced current, \Eq{A_equation} is modified to
\begin{align}
\Ddot{\mathcal{A}}_{\pm} +(\sigma+H)\dot{\mathcal{A}}_{\pm}+\frac{k}{a}
\left(\frac{k}{a}\mp g_{\phi\gamma}\dot{\phi}\right)\mathcal{A}_{\pm}=0.
\label{mod_A_equation}
\end{align}
Here, the conductivity $\sigma$ is given by
\begin{align}
    \sigma=\frac{e^3}{6 \pi^2 H} |B| \coth\left(\frac{\pi |B|}{E}\right) e^{-\frac{\pi m^2}{eE}}
\end{align}
Here, $|B|$ denotes the magnitude of the magnetic field component parallel to the electric field, $E$. To evaluate the conductivity, we use $E=\sqrt{2\rho_E}$, $B=\sqrt{|\rho_{\EE\cdot \BB}|/(2 \rho_E)}$, where $\rho_E$ is defined in \Eq{energy_densities} and $\rho_{\EE\cdot \BB}$ is given by,
\begin{align}
\rho_{\EE\cdot \BB}=\frac{1}{a^4} \int d \ln k \frac{k^4}{2\pi^2 }(|\mathcal{A}_+||\dot{\mathcal{A}}_+|-|\mathcal{A}_-||\dot{\mathcal{A}}_-|).
\end{align}
We solve \Eq{mod_A_equation} numerically using the \texttt{axionU1back.f90} module of the Pencil Code. In the absence of a significant Schwinger current, the time step is determined by requiring that the inverse time step resolve the largest relevant dynamical scale,
\begin{align}
    \Delta t^{-1}
    \gtrsim
    \max\left(H,m_\phi,\frac{k}{a}\right).
\end{align}
After including the charged current produced via Schwinger mechanism, the time-step criterion modifies to,
\begin{align}
    \Delta t^{-1}
    \gtrsim
    \max\left(H,m_\phi,\frac{k}{a},\sigma\right).
\end{align}

Here, a numerical difficulty arises because the conductivity can rapidly become much larger than all the other relevant dynamical scales, namely $H$, $m_\phi$, and $k/a$. Consequently, resolving the time scale $\sigma^{-1}$ requires an extremely small time step, making the numerical evolution for O(10) e-folds computationally expensive. To overcome this difficulty, we implement a switching procedure between the full second-order evolution equation and its overdamped limit. When the damping term becomes dominant, such that
\begin{align}
\sigma \gg\sqrt{\left| \frac{k}{a} \left( \frac{k}{a}\mp g_{\phi\gamma}\dot{\phi}\right)\right|},
\end{align}
the system enters an overdamped regime. In this case, the
$\ddot{\mathcal{A}}$
can be neglected, and \Eq{mod_A_equation} reduces to the first-order equation
\begin{align}
\dot{\mathcal{A}}_{\pm}=-\frac{1}{\sigma+H}\frac{k}{a} \left(\frac{k}{a}\mp g_{\phi\gamma}\dot{\phi}\right)\mathcal{A}_{\pm}. \label{overdamped_A_equation}
\end{align}
This approximation allows the evolution to be followed without resolving the much shorter time scale $\sigma^{-1}$. During the subsequent evolution, the conductivity may decrease and the system can leave the overdamped regime. Whenever the damping term is no longer sufficiently large compared with the characteristic frequency of the gauge-field mode, we switch back to solving the full second-order equation, \Eq{mod_A_equation}. This switching procedure significantly improves the numerical efficiency while retaining the full dynamical evolution whenever the overdamped approximation is no longer valid.
\begin{table}
\begin{center}
\begin{tabular}{p{.6cm}|p{1.8cm}|p{1.8cm}|p{1.8cm}|p{1.8cm}|p{0.6cm}|p{.6cm}|p{.7cm}}
\hline
Run & $g_{\phi \gamma}$ (in~$m_{\rm pl}^{-1}$) &$\Lambda$ (in GeV)& $\frac{\Lambda^2}{\phi_0} $ (in GeV) & grid size& $N_{e}$  &$n_{min}$&$n_{max}$ \\
\hline
A1 & $400$ & $100$ & $4.1 \times 10^{-16}$     & $256$ & $2.87$  &$4$ &$8$\\ 
A1s & $400$ & $100$ & $4.1 \times 10^{-16}$     & $256$ & $2.87$  &$4$ &$8$\\ 
A4s & $800$ & $100$ & $4.1 \times 10^{-16}$     & $256$ & $2.87$  &$4$ &$9$\\ 
A5s & $1600$ & $100$ & $4.1 \times 10^{-16}$     & $256$ & $2.87$  &$4$ &$10$\\ 
A6s & $20000$ & $100$ & $4.1 \times 10^{-16}$     & $256$ & $0.27$  &$4$ &$10$\\ 
B & $1200$ & $1$ & $4.1 \times 10^{-18}$     & $256$ & $2.87$  &$4$ &$8$\\ 
C & $1200$ & $10$ & $4.1 \times 10^{-17}$     & $256$ & $2.87$  &$4$ &$9$\\ 
D & $400$ & $10000$ & $4.1 \times 10^{-14}$     & $256$ & $2.87$  &$4$ &$8$\\ 
\hline
\end{tabular}
\caption{The model parameter for run A1 (without Schwinger current) and for runs A1s, B, C and D (with Schwinger current) for different scale of GMSSM inflation ($\Lambda$).}\label{table3}
\end{center}
\end{table}
\subsection{Gauge field production in the GMSSM with Schwinger currents}
\begin{figure}
    \centering
    \includegraphics[width=1\linewidth]{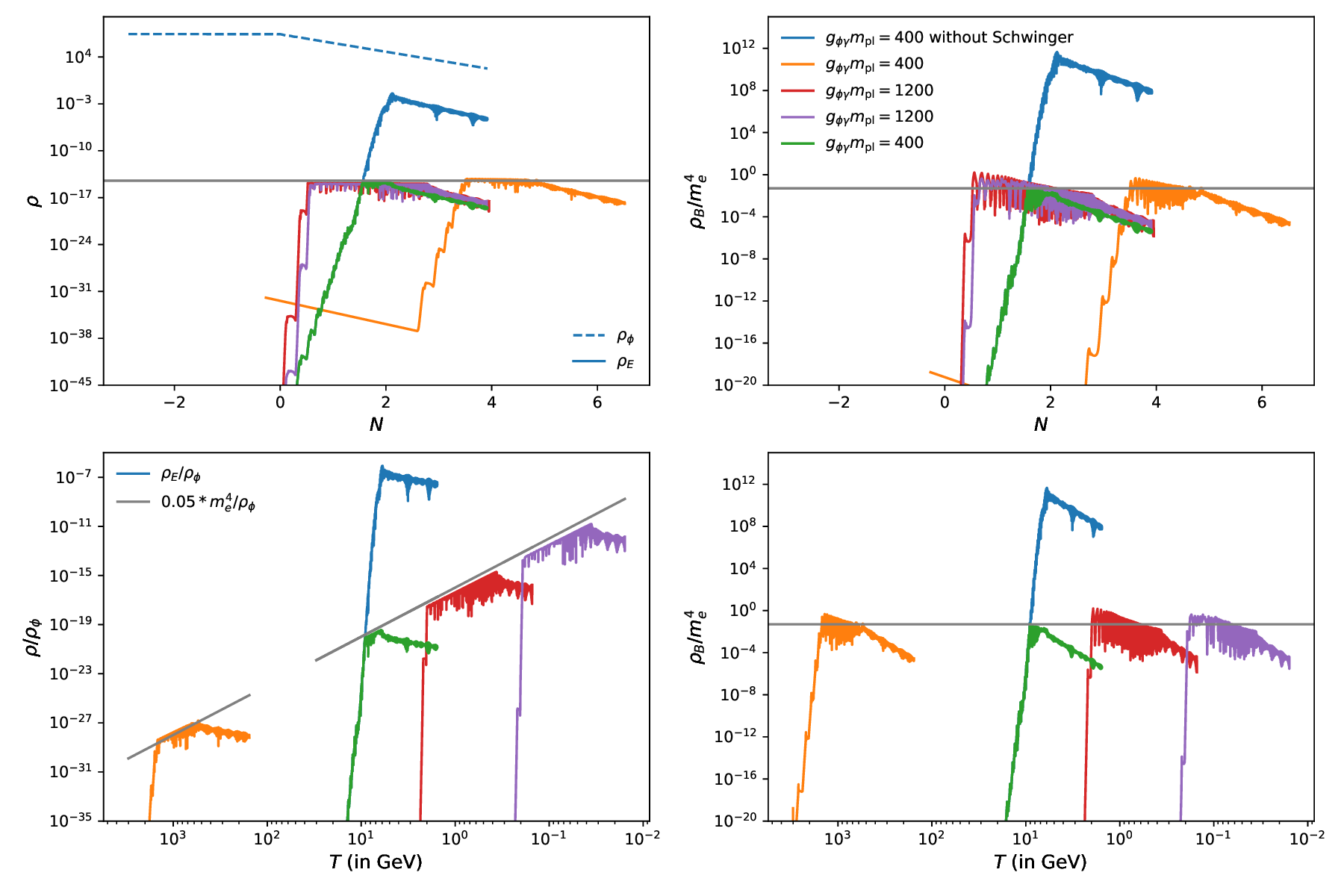}
    \caption{Evolution of the inflaton and gauge field energy densities as functions of the number of e-folds, $N$ (top panel), and the temperature, $T$ (lower panel), for different values of the coupling strength $g_{\phi\gamma}$ and the inflationary energy scale $\Lambda$. The parameters used in the simulations shown in this figure are given in \Tab{table3}. The horizontal grey line corresponds to a value $0.05~m_e^4$ in the top left panel and $0.05$ in the top right panel.}
    \label{gmssm_cond}
\end{figure}

Using the strategy described in the previous section, we solve \Eq{mod_A_equation} numerically for different values of the inflationary energy scale, denoted by $\Lambda$, in the GMSSM inflation model. The model parameters used in our simulations are listed in \Tab{table3}. The resulting energy densities are shown in \Fig{gmssm_cond}. The top left panel shows the evolution of the inflaton energy density, $\rho_{\phi}$, and the electric part of the gauge field energy density, $\rho_E$, as functions of the number of e-folds, $N$. The top right panel shows the evolution of the energy density of the magnetic part of the gauge field, $\rho_B$, normalized by $m_e^4$. The lower left panel shows the evolution of the $\rho_E$ normalized by $\rho_{\phi}$, as a function of the temperature, $T\equiv (30\rho_{\phi}/(g_r\pi^2))^{1/4}$, where $g_r$ denotes the number of relativistic degrees of freedom contributing to the energy density. This temperature, $T$ corresponds to the temperature that the Universe would have if reheating occurred instantaneously at that time.
For simplicity, we take $g_r=100$ throughout our analysis. The lower right panel shows the evolution of $\rho_B/m_e^4$ as a function of $T$. In each panel, the blue and green curves correspond to the case with $\Lambda=100$ GeV and $g_{\phi\gamma}=400 m_{\rm pl}^{-1}$. The blue curves represent the results without including the Schwinger current, whereas the green curves include the effects of the Schwinger-induced current. The orange curve corresponds to a higher inflationary energy scale, $\Lambda=10^4$ GeV, while the red and purple curves correspond to lower energy scales, $\Lambda=10$ GeV and $\Lambda=1$ GeV, respectively. For the red and purple curves, we use a larger coupling, $g_{\phi\gamma}=1200\,m_{\rm pl}^{-1}$.

From the comparison between the blue and green curves, we find that the gauge field energy density saturates at approximately $\rho_E+\rho_B\sim 0.05~m_e^4$ after the Schwinger current is included. This saturation occurs because, once the gauge field strength becomes sufficiently large, the Schwinger induced conductivity becomes large enough to compensate for the continued production of gauge fields through their coupling to the inflaton. The Schwinger current becomes important only when $\rho_E$ becomes comparable to or larger than $m_e^4$, since the exponential suppression in \Eq{cond_exp} then becomes negligible.
Furthermore, we observe that $\rho_B$ remains approximately of order $m_e^4$ for some duration before eventually decreasing due to the usual Hubble dilution, $\rho_B\propto a^{-4}$. We also find that for larger values of $g_{\phi\gamma}$, $\rho_B$ remains of order $m_e^4$ for a longer period before transitioning to the decays via Hubble dilution.

\begin{figure}
    \centering
    \includegraphics[width=1\linewidth]{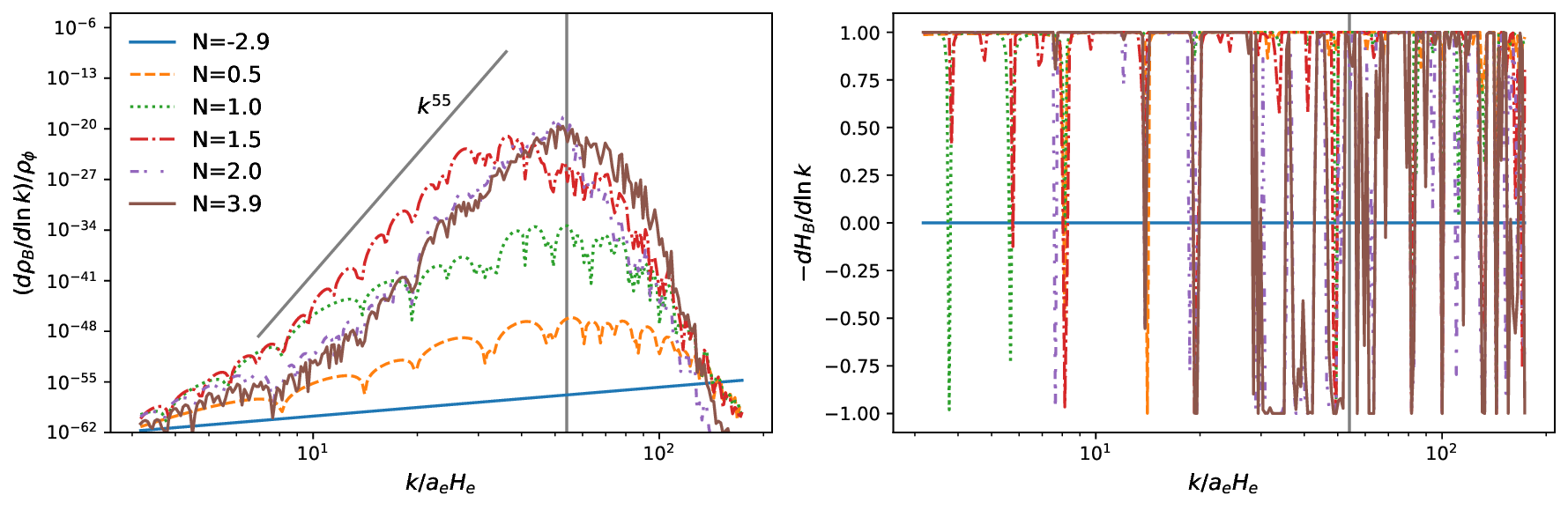}
    \caption{Gauge field energy density spectra at $N=-2.9$, $0.5$, $1.0$, $1.5$, $2.0$, $3.9$ for the A1s run. The right panel show the corresponding helicity spectra defined in \Eq{hel_spectra}.}
    \label{spec_gmssm_cond}
\end{figure}

\begin{figure}
    \centering
    \includegraphics[width=1\linewidth]{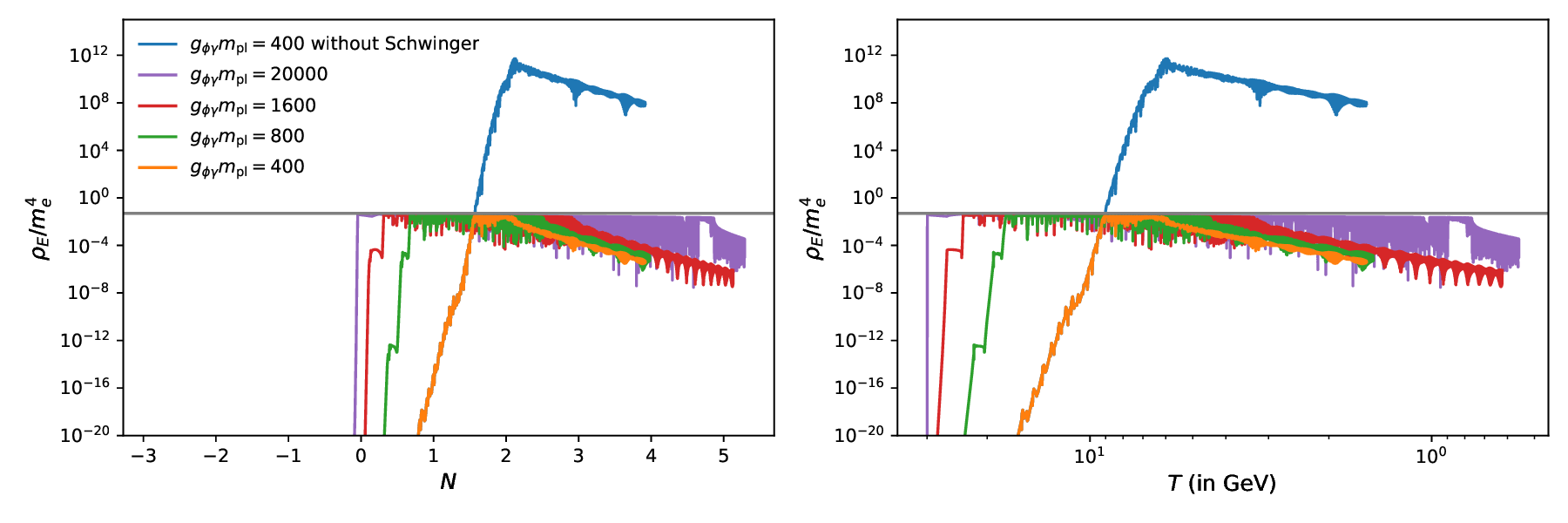}
    \caption{Evolution of the energy density of the electric part of the gauge field normalized by $m_e^4$ vs $N$ (left panel) and $T$ (right panel), for different values of the coupling $g_{\phi\gamma}$. The plots correspond to the A1(blue curve), A1s (yellow curve), A4s(green curve) and A5s (red curve) runs, with $\Lambda=100$ GeV. The horizontal grey line corresponds to a value $0.05~m_e^4$ in the left panel and $0.05$ in the right panel.}
    \label{two_panel_different_g}
\end{figure}

Furthermore, we plot the spectra of the magnetic field energy density in \Fig{spec_gmssm_cond} at different times, corresponding to $N=-2.9$ (blue curve), $0.5$ (orange curve), $1.0$ (green curve), $1.5$ (red curve) and $2.0$ (purple curve) and $3.9$ (brown curve). Comparing these spectra with those shown in \Fig{spec_gmssm} for runs with similar parameters but without the Schwinger current, we find that the inclusion of the Schwinger current leads to a suppression of power in all the wave-modes and the peak of the spectrum remains at the same wavenumber, $k/(a_e~H_e)=54$.
From this numerical analysis, we conclude that the maximum gauge-field energy density reached during low-scale inflation is of order $m_e^4$. The gauge-field energy density remains approximately of order $m_e^4$ for a certain period following its initial growth, before eventually decreasing as $a^{-4}$ during the post-inflationary phase. In \Fig{two_panel_different_g}, we show the evolution of the electric part of the gauge field energy density as a function of $N$ (left panel) and as a function of the
temperature $T$ (right panel)
for different values of the coupling strength $g_{\phi\gamma}$. The results correspond to runs A1 (blue curve), A1s (orange curve), A4s (green curve), A5s (red curve) and A6s (purple curve), with an inflationary energy scale of $\Lambda=100~{\rm GeV}$. From this figure, we can conclude that the duration over which the gauge-field energy density remains of order $m_e^4$ increases with the coupling strength $g_{\phi\gamma}$. Therefore, by increasing $g_{\phi\gamma}$, this period can be extended such that the gauge-field energy density remains approximately of order $m_e^4$ until the end of the post-inflationary phase. Consequently, the gauge-field energy density at the onset of the radiation-dominated era can be taken of order $m_e^4$.

\section{Magnetogenesis in Low-scale inflation}\label{section4}
If the gauge field considered in this paper is identified with the Standard Model U(1) field, it may explain the origin of the observed large-scale magnetic fields in the Universe. Here, we investigate the possibility of magnetogenesis in a low-scale inflationary scenario. The motivation for exploring low-scale inflation is twofold. First, magnetic fields produced in high-scale inflation are typically too small to satisfy the lower bounds inferred from $\gamma$-ray observations of blazars \cite{Iarygina:2025ncl,Domcke:2018eki,Gorbar:2021zlr}. Second, magnetogenesis occurring at low inflationary scales naturally avoids the overproduction of baryon isocurvature perturbations \cite{Kamada:2020bmb}. This latter constraint is perhaps less reliable, as
recent studies which have revisited the baryon-isocurvature constraints, find that they do not necessarily rule out the generation of primordial magnetic fields before the EW era \cite{Boyer:2025jno,Hamada:2025ooy}.

\begin{figure}
    \centering
    \includegraphics[width=1\linewidth]{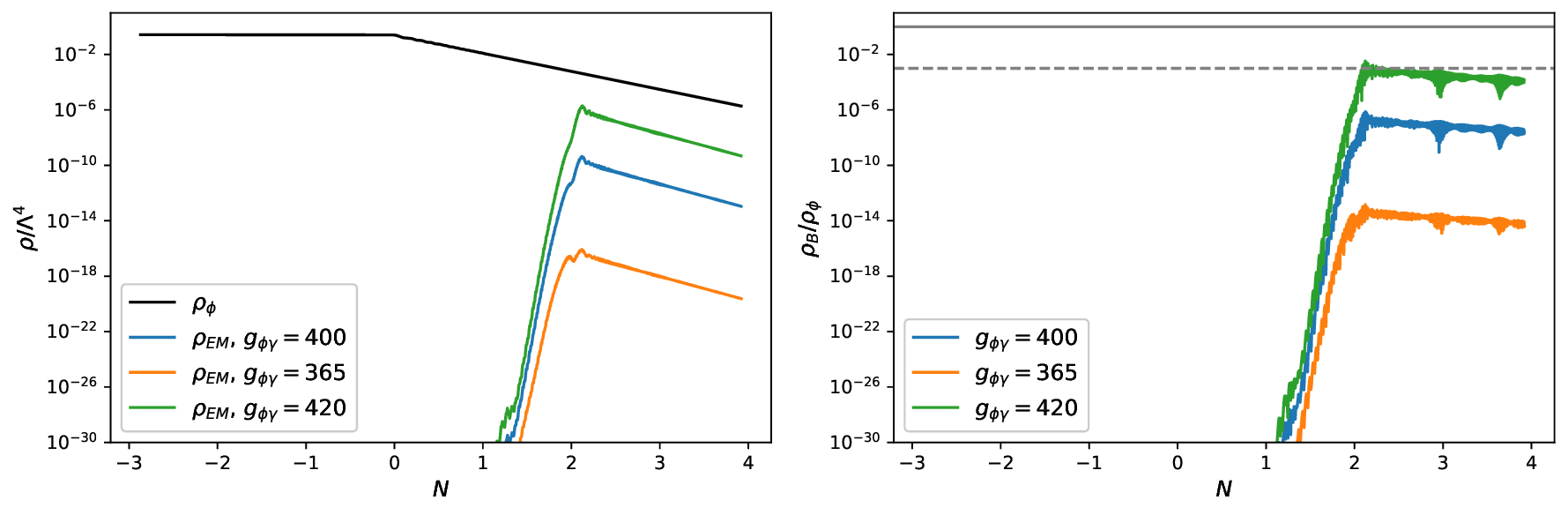}
    \caption{The evolution of the $\rho_{\phi}$ and $\rho_{EM}$ for different $g_{\phi\gamma}$ is shown in the left panel. In the right panel, we show the ratio of the magnetic component of the gauge field energy density to $\rho_{\phi}$. The solid and dashed horizontal grey line in the right panel corresponds to a value 1 and $10^{-3}$, respectively.}
    \label{saturation_energy_density}
\end{figure}

Let us first consider the case when the effects of the Schwinger current is not included. As shown in \Fig{saturation_energy_density}, the rate of the gauge field amplification increases with the coupling strength, $g_{\phi \gamma}$. And the produced gauge field strength can reach up to a significant fraction of the total energy density for $g_{\phi \gamma}\sim O(400) m_{\rm pl}^{-1}$ as shown in the right panel of \Fig{saturation_energy_density} in the case when the effects of the charged current via Schwinger mechanism is not considered.
As evident from \Fig{spec_gmssm}, the gauge field spectrum has a broken power law form, peaking at a wavenumber corresponding to $k/(a_eH_e)=54$. To estimate the present-day magnetic field strength, we assume that reheating occurs $N_r$ e-folds after the gauge field reaches saturation. After reheating, the Universe is filled with a highly conducting charged plasma, with plasma conductivity much larger than the Hubble parameter, and subsequently evolves in a radiation dominated phase. In such a medium, the electric component of the gauge field is rapidly damped, while the magnetic component evolves according to freely decaying magnetohydrodynamics, in addition to dilution from cosmic expansion. The nonlinear evolution of magnetic fields has been studied extensively, both analytically and numerically \cite{Banerjee:2004df,Subramanian:2015lua,Brandenburg:2016odr,Brandenburg:2017neh,Zhou:2019hkc,Bhat:2020anu,Hosking:2020wom,Zhou:2022xhk,2024OJAp....7E..75D,Kasselmann:2026tiu}.

To estimate the present-day magnetic field strength and coherence length, accounting for their post-generation evolution, we use the expressions given in Ref.~\cite{Sharma:2024nfu}.
\begin{align}\label{mfstrength}
B_\mathrm{rms}\big|_0
&=9.2 \times 10^{-15}\, {\rm G}\,\sqrt{\frac{\rho_B}{\rho_{\phi}}\Big|_s}  \left(\frac{ 10^{-6} m_{\rm pl}}{H_s}\right)^{1/6}r^{1/3}
\end{align}and 
\begin{align}\label{cohlen}
L_c^\mathrm{NL}=0.8~ {\rm pc} ~(\mathcal{H}L_c)\left(\frac{10^{-6}m_{\rm pl}}{H_s}\right)^{1/6}\,r^{-2/3}.
\end{align}
Here $r=\max(1,\mathcal{H}~L_c/V_A)$, where $V_A\equiv\sqrt{B^2/(\rho+p)}$ is the Alfv\'en velocity. $\rho_B/\rho_{\phi}\big|s$ denotes the ratio of the magnetic energy density to that of the $\phi$ field at the time of saturation, while $H_s$ is the Hubble parameter at saturation. Furthermore, $\mathcal{H}_s L_c$ represents the coherence length in units of the comoving Hubble radius at saturation, which in our case depends on $g_{\phi \gamma} m_{\rm pl}$. 

\begin{figure}
    \centering
    \includegraphics[width=1\linewidth]{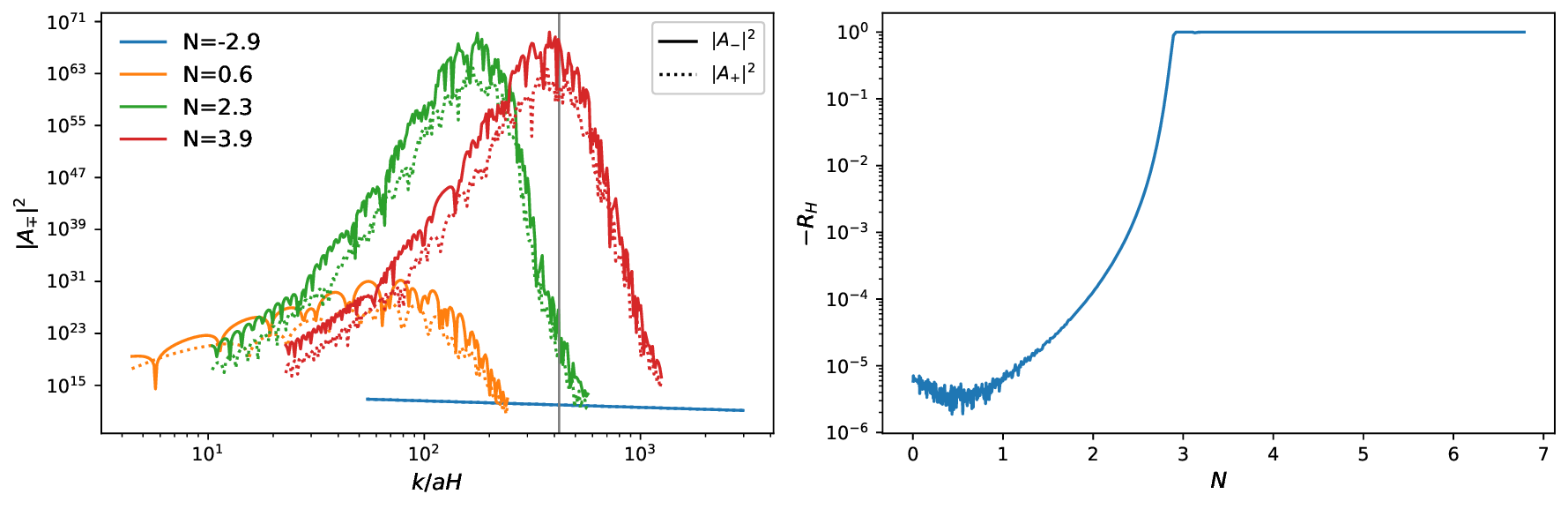}
    \caption{$\left|A_{-}\right|^2$ and $\left|A_{+}\right|^2$ vs $k$ at different times, corresponding to $N=-2.9$ (blue curves), $0.6$ (yellow curves), $2.3$ (green curves), and $3.9$ (red curves) is shown in the left panel. Solid and dashed curves denote the negative- and positive-helicity modes, respectively. Time evolution of the quantity $R_H$, defined in \Eq{def_rh} is shown in the right panel. Both panels show results for the A1 run.}
    \label{helicity_est}
\end{figure}

In estimating the magnetic field strength and its coherence length at the present epoch, we have assumed that the primordial magnetic field is fully helical. As can be seen from \Fig{spec_gmssm}, the helicity fraction is close to $-1$ for most wavenumbers, while it approaches $1$ for some modes, leading to oscillatory features in the spectrum. To demonstrate this further, we show the comparison between the spectra of the negative and positive helicity modes in the left panel of \Fig{helicity_est} for the A1 run. The figure shows that, for most of the evolution after the onset of gauge field amplification, the negative helicity mode dominates over the positive helicity mode for the majority of wavenumbers. To obtain a wavenumber independent estimate of the helicity, we further define the quantity $R_H$ as
\begin{align}\label{def_rh}
R_H=\frac{\int d \ln k k^5 (|\mathcal{A}_+|^2-|\mathcal{A}_-|^2)}{\int d \ln k k^5 (|\mathcal{A}_+|^2+|\mathcal{A}_-|^2)}
\end{align}
and its evolution with time is shown in the right panel of \Fig{helicity_est}. As evident from \Fig{helicity_est}, $R_H$ approaches negative unity as expected for a fully helical field.
In summary, in estimating the subsequent evolution of the magnetic field, we assume the evolution appropriate for a fully helical field. A precise treatment of the evolution of such a spectrum would require full three-dimensional magnetohydrodynamic simulations. Nevertheless, we expect that this approximation does not significantly affect our estimates of the present day magnetic field strength.

\begin{table}
\begin{center}
\begin{tabular}{p{.6cm}|p{1.25cm}|p{1.8cm}|p{1.8cm}|p{1.8cm}|p{1.8cm}|p{1.8cm}}
\hline
Run & $g_{\phi \gamma}$ (in~${m_\mathrm{pl}^{-1}}$) &$H_s/m_{\rm pl}$ &$\rho_B/\rho_{\phi}|_s$& $B_{rms}|_0$ (in G)& $L_c^{NL}$ (in kpc) & $B_\mathrm{bound}$ (in G) \\
\hline
A1 & $400$ & $4.0 \times 10^{-36}$ &$7.5 \times 10^{-7}$& $9.0 \times 10^{-13}$         & $9.9 \times 10^{-2}$& $8.1 \times 10^{-16}$\\ 
A2 & $365$ & $4.0 \times 10^{-36}$ &$1.5 \times 10^{-13}$& $2.0 \times 10^{-14}$         & $5.9 \times 10^{-4}$& $1.0 \times 10^{-14}$\\ 
A3 & $420$ & $4.0 \times 10^{-36}$ &$3.3 \times 10^{-3}$& $4.2 \times 10^{-11}$         & $1.5 \times 10^{-1}$& $6.5 \times 10^{-16}$\\ 

\hline
\end{tabular}
\caption{Estimates of the present-day magnetic field strength ($B_{\rm rms}|0$) and its coherence length ($L_c^{\rm NL}$) for different values of $g_{\phi \gamma}$ are given. Here, $H_s$ and$\rho_B/\rho_{\phi}\big|s$ denote the Hubble parameter and the ratio of the magnetic energy density to that of the $\phi$ field at the time of saturation, respectively. The quantity $B_{\rm bound}$ represents the lower bound on the magnetic field strength inferred from the non-observation of GeV photons in blazar observations. To obtain this bound, we use $B_{\rm bound}=1.8 \times 10^{-17} G (L/0.3 {\rm Mpc})^{-1/2}$ \cite{MAGIC:2022piy}. These estimates are for the case in which the effect of charged currents produced via the Schwinger mechanism has not been considered.}\label{table2}
\end{center}
\end{table}

Our estimates of the magnetic field and its coherence length obtained using \Eq{mfstrength} and \Eq{cohlen} are given in \Tab{table2}. For $g_{\phi\gamma}\sim 420 {\rm m_{pl}^{-1}}$, the produced gauge field energy density reached up to 0.1\% of the total energy budget. For this case, the obtained magnetic field strength incorporating the late time non-linear processing of the magnetic field, we obtained the present day value of the magnetic field is $4.2 \times 10^{-11}~{\rm G}$ and its coherence length is equal to $0.15~{\rm kpc}$.

\begin{figure}
    \centering
    \includegraphics[width=1\linewidth]{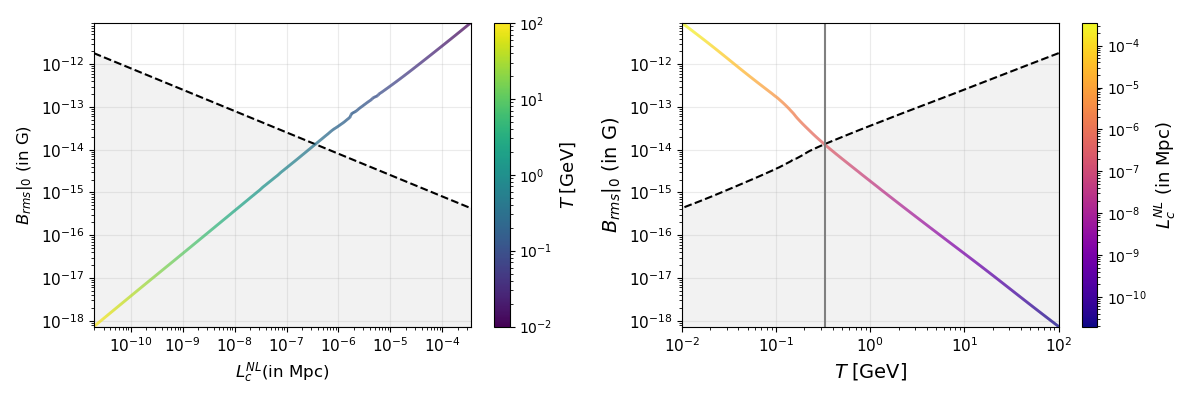}
    \caption{Present-day magnetic-field strength, $B_{rms}|_0$ and coherence length, $L_c^{NL}$ for different reheating temperatures, $T$. The left panel shows $B_{rms}|_0$ versus $L_c^{NL}$ with colors on the solid line indicating the reheating temperature, while the right panel shows $B_{rms}|_0$ versus T, with colors on the solid line indicating the corresponding coherence length. The dashed black line denotes the lower bound from blazar observations \cite{MAGIC:2022piy}. The gray-shaded region represents the parameter space excluded by the lower bound derived from the non-detection of GeV photons from blazars \cite{MAGIC:2022piy}. In obtaining these estimates, we assume that the magnetic field energy density at the end of reheating is $0.05~m_e^4$, where $m_e$ is the electron mass. The vertical grey line in the right panel corresponds to a value, $T=0.33$ GeV.
}
    \label{mf_bound}
\end{figure}

As discussed in the previous section, after including the charged current generated through the Schwinger mechanism, the gauge-field energy density initially increases during the post-inflationary phase and subsequently reaches an amplitude of order $m_e^4$. After remaining approximately at this level for some duration, it begins to decrease as $a^{-4}$. We also showed that the coupling strength $g_{\phi\gamma}$ can be tuned such that the gauge field energy density remains of order $m_e^4$ until the end of the post-inflationary phase, corresponding approximately to the onset of radiation domination. We further estimate the present day magnetic field strength and coherence length for different reheating temperatures using \Eqs{mfstrength}{cohlen}. In these estimates, we assume that the magnetic-field energy density at the end of reheating is $0.05~m_e^4$ and that the ratio of the coherence length to the Hubble horizon is
approximately $400$. The resulting magnetic field strength and coherence length are shown in \Fig{mf_bound}. The left panel shows the magnetic-field strength as a function of its coherence length, with different colors corresponding to different reheating temperatures. In the right panel, we show the magnetic-field strength as a function of the reheating temperature, with different colors along each curve indicating the corresponding coherence length. The dashed black line denotes the lower bound inferred from the non-detection of GeV photons from blazar observations,
$B_{\rm bound}=1.8\times10^{-17} (L_C^{NL}/0.2{\rm Mpc})^{-1/2} {\rm G}$ \cite{MAGIC:2022piy}. As evident from the right panel of \Fig{mf_bound}, explaining the presence of intergalactic magnetic fields \cite{Neronov:1900zz,Taylor:2011bn,MAGIC:2022piy} requires reheating to occur at a temperature below $\sim 0.3~{\rm GeV}$. Assuming that the magnetic field energy density is $0.05~m_e^4$ until the reheating temperature of around $100$ MeV, the corresponding magnetic-field strength and coherence length at the present epoch are $2\times10^{-13} {\rm G}$ and $5\times10^{-6} {\rm Mpc}$, respectively. As shown in the right panel of \Fig{two_panel_different_g}, for an inflationary energy scale of $\Lambda=100~{ \rm GeV}$, a coupling strength $g_{\phi\gamma}>2\times10^4\,m_{\rm pl}^{-1}$ is required to sustain the saturation regime until the reheating temperature less than $1~{\rm GeV}$. This can also be achieved for higher inflationary energy scales; doing so would require an even larger coupling strength.

In our estimates of the present day magnetic field strength and its coherence length, we have taken the Alfv\'en crossing time ($\tau_A$) as the relevant timescale. However, recent studies beginning with 
\cite{Zhou:2019hkc,Bhat:2020anu} and then \cite{Hosking:2020wom,Hosking:2022umv} have suggested that the magnetic reconnection time ($\tau_{\rm rec}$) is more appropriate for describing the turbulent evolution of primordial magnetic fields. The reconnection time is longer than the Alf\'ven crossing time, and its dependence on the magnetic Prandtl number remains unclear \cite{Hosking:2022umv,Brandenburg:2024tyi,2026arXiv260117955A}. If we take the reconnection time as the relevant timescale, the estimated magnetic field strength increases by a factor of $(\tau_{\rm rec}/\tau_A)^{1/3}$, while the coherence length decreases by a factor of $(\tau_{\rm rec}/\tau_A)^{-2/3}$. Nevertheless, since the evolution of a maximally helical magnetic field conserves $B^2L$, this change in the timescale does not change our conclusion regarding whether the resulting magnetic fields can satisfy the blazar lower bound.
\section{Discussion and conclusion}\label{section5}
In this paper, we study the generation and dynamics of gauge fields in a low scale inflation model for the case in which the gauge field interacts with the inflaton through a coupling of the form $\phi F_{\mu \nu} \tilde{F}^{\mu\nu}$. As a representative scenario, we considered the GMSSM inflation model. For the coupling strengths of interest in this paper the back
reaction of the produced gauge fields on the inflaton dynamics is not important. In the low-scale inflation model, the gauge fields are only amplified post-inflation. This is because in low-scale inflation the inflaton velocity is small, suppressing the tachyonic growth during inflation.

In particular, we consider inflation occurring around the EW phase transition scale. We find that, for a coupling strength of $g_{\phi\gamma}\simeq 420~m_{\rm pl}^{-1}$, the magnetic field energy density reaches approximately $0.1\%$ of the background inflaton energy density. 
After accounting for the subsequent nonlinear evolution of the generated magnetic field, this corresponds to a present-day magnetic field strength of $\sim 4 \times 10^{-11}~{\rm G}$ with a coherence length of $\sim 0.15~{\rm kpc}$. 
This combination of magnetic field strength and coherence length lies well above the lower bound inferred from the non-detection of GeV photons in blazar observations.

However, the amplified gauge fields can induce the production of charged particles through the Schwinger mechanism. The resulting charged current modifies the gauge field evolution and, consequently, the estimates obtained when this effect is neglected. The impact of the Schwinger induced current on gauge field production has been studied extensively in the literature \cite{Kobayashi:2014zza,Domcke:2018eki,Sobol:2019xls,Gorbar:2021zlr,Iarygina:2025ncl}, where it has been shown that the backreaction of the induced current can significantly suppress gauge field amplification. Most of these studies, however, have focused on high scale inflationary scenarios, typically 
with inflation occurring near the GUT scale.
In the present work, we are primarily interested in magnetogenesis during low scale inflation. In this regime, the dominant amplification of the gauge field occurs during the post inflationary phase, when the inflaton undergoes coherent oscillations, as shown in \Fig{gmssm_plot}. We therefore investigate the effect of the Schwinger induced charged current on gauge-field amplification during this phase.

We find that the induced current suppresses gauge field amplification and causes the gauge field energy density to saturate at a value of order $m_e^4$. This behaviour can be understood from the exponential dependence of the Schwinger current on the electric field strength, as shown in \Eq{cond_exp}. When the electric field strength is below $m_e^2$, charged particle production is exponentially suppressed, and the effect of the induced current is negligible. Once the gauge field strength becomes comparable to or larger than $m_e^2$, charged-particle production becomes efficient, and the resulting current provides increasingly strong damping of the gauge field. Further amplification therefore requires a sufficiently large coupling $g_{\phi\gamma}$ to overcome this suppression, resulting in the saturation of the gauge field energy density at approximately $m_e^4$. We also find that, although both helicity modes of the gauge field are amplified post inflation, one helicity dominates the total energy density. This occurs because one helicity experiences additional amplification just before the scalar field begins to oscillate, and this dominance continues during the later evolution.

The Hubble scale during the inflationary epoch considered in our analysis is smaller than $m_e$. We further find that the coupling strength $g_{\phi\gamma}$ can be tuned such that the gauge field energy density remains approximately of order $m_e^4$ until the end of the post inflationary era. Assuming a magnetic field energy density of $0.05~m_e^4$ at the end of this era, we obtain a present day magnetic field strength of approximately $2\times10^{-13}~{\rm G}$ and a coherence length of $5\times10^{-6}~{\rm Mpc}$ for a reheating temperature of $100~{\rm MeV}$. These values are compatible with the lower bound on intergalactic magnetic fields inferred from the non-detection of secondary GeV photons from blazar observations. Furthermore, we find that obtaining a magnetic field strength above this lower bound requires the post inflationary era to end at a temperature below approximately $0.3~{\rm GeV}$. For an inflationary energy scale of $\Lambda=100~{\rm GeV}$, the coupling strength should be greater than $2\times10^4~m_{\rm pl}^{-1}$ to sustain the saturation of the gauge-field energy density for the reheating temperature below $1~{\rm GeV}$. This condition can also be satisfied for higher inflationary energy scales, but it requires an even larger coupling strength. Furthermore, we find that the magnetic field energy density spectrum obtained at the end of magnetogenesis is significantly steeper than that produced in high scale inflationary scenarios. During its subsequent evolution in the radiation dominated Universe, the spectrum is expected to become considerably shallower on length scales for which the eddy turnover time is shorter than the Hubble time.

Magnetogenesis in the GMSSM model has also been investigated in Ref.~\cite{Yanagihara:2023qvx} but without taking account of the Schwinger current. Although we use the GMSSM model as a representative example of low-scale inflation, the main conclusions of this work are expected to remain qualitatively valid for other low-scale inflationary models with a similar inflationary energy scale. The reason is that the amplification of the gauge field is primarily governed by the term $g_{\phi\gamma}\dot{\phi}$ in the gauge field equation of motion. After inflation, the inflaton begins coherent oscillations around the minimum of its potential once the Hubble parameter becomes comparable to its effective mass, $H\sim m_{\rm eff} =\sqrt{d^2V(\phi)/d \phi^2}$. During this stage, the typical magnitude of the inflaton velocity is set by the inflationary energy scale and is generally of order $\dot{\phi}\sim H m_{\rm Pl}$, up to model-dependent factors of order unity. Consequently, for a fixed inflationary energy scale and comparable values of $g_{\phi\gamma}$, the amplification of the gauge field is expected to be of a similar magnitude for other low scale inflationary models. Quantitative differences may nevertheless arise from the detailed shape of the inflaton potential and the resulting oscillation dynamics.

We note that post-inflationary amplification has not been considered in earlier high-scale (GUT-scale) studies, which also reported that it is not possible to generate sufficiently strong magnetic fields to explain the lower bound in the intergalactic regime suggested by the non-observations of the GeV photons in blazars \cite{Iarygina:2025ncl}. Incorporating the post-inflationary amplification mechanism into high-scale axion-U(1) models may change the conclusions, and will thus be interesting to study.

\acknowledgments

RS would like to thank the Department of Science and Technology, India, for financial support through the INSPIRE Faculty Fellowship (DST/INSPIRE/IFFCALL-2024/IFA-24-PH-313). DK and SS acknowledge funding from the DST-SERB/ANRF Power Grant no. SPG/2021/002228 for computational facility setup. KS was partially supported by the Alexander Von Humboldt Foundation through the
Carl Friedrich von Siemens Research Award. He thanks Volker Springel for warm hospitality
at the Max-Planck Institute for Astrophysics, Garching.

\bibliographystyle{JHEP}
\bibliography{references}

\providecommand{\href}[2]{#2}\begingroup\raggedright\begin{thebibliography}{10}

\bibitem{Widrow:2002ud}
L.M.~Widrow, \emph{{Origin of galactic and extragalactic magnetic fields}}, \href{https://doi.org/10.1103/RevModPhys.74.775}{\emph{Rev. Mod. Phys.} {\bfseries 74} (2002) 775} [\href{https://arxiv.org/abs/astro-ph/0207240}{{\ttfamily astro-ph/0207240}}].

\bibitem{Bernet:2008qp}
M.L.~Bernet, F.~Miniati, S.J.~Lilly, P.P.~Kronberg and M.~Dessauges-Zavadsky, \emph{{Strong magnetic fields in normal galaxies at high redshifts}}, \href{https://doi.org/10.1038/nature07105}{\emph{Nature} {\bfseries 454} (2008) 302} [\href{https://arxiv.org/abs/0807.3347}{{\ttfamily 0807.3347}}].

\bibitem{Beck:2008ty}
R.~Beck, \emph{{Galactic and Extragalactic Magnetic Fields}}, \href{https://doi.org/10.1063/1.3076806}{\emph{AIP Conf. Proc.} {\bfseries 1085} (2009) 83} [\href{https://arxiv.org/abs/0810.2923}{{\ttfamily 0810.2923}}].

\bibitem{Kronberg:2007dy}
P.P.~Kronberg, M.L.~Bernet, F.~Miniati, S.J.~Lilly, M.B.~Short and D.M.~Higdon, \emph{{A Global Probe of Cosmic Magnetic Fields to High Redshifts}}, \href{https://doi.org/10.1086/527281}{\emph{Astrophys. J.} {\bfseries 676} (2008) 7079} [\href{https://arxiv.org/abs/0712.0435}{{\ttfamily 0712.0435}}].

\bibitem{Neronov:1900zz}
A.~Neronov and I.~Vovk, \emph{{Evidence for strong extragalactic magnetic fields from Fermi observations of TeV blazars}}, \href{https://doi.org/10.1126/science.1184192}{\emph{Science} {\bfseries 328} (2010) 73} [\href{https://arxiv.org/abs/1006.3504}{{\ttfamily 1006.3504}}].

\bibitem{Tavecchio:2010mk}
F.~Tavecchio, G.~Ghisellini, L.~Foschini, G.~Bonnoli, G.~Ghirlanda and P.~Coppi, \emph{{The intergalactic magnetic field constrained by Fermi/LAT observations of the TeV blazar 1ES 0229+200}}, \href{https://doi.org/10.1111/j.1745-3933.2010.00884.x}{\emph{Mon. Not. Roy. Astron. Soc.} {\bfseries 406} (2010) L70} [\href{https://arxiv.org/abs/1004.1329}{{\ttfamily 1004.1329}}].

\bibitem{Dolag:2010ni}
K.~Dolag, M.~Kachelriess, S.~Ostapchenko and R.~Tomas, \emph{{Lower limit on the strength and filling factor of extragalactic magnetic fields}}, \href{https://doi.org/10.1088/2041-8205/727/1/L4}{\emph{Astrophys. J. Lett.} {\bfseries 727} (2011) L4} [\href{https://arxiv.org/abs/1009.1782}{{\ttfamily 1009.1782}}].

\bibitem{Essey:2010nd}
W.~Essey, S.~Ando and A.~Kusenko, \emph{{Determination of intergalactic magnetic fields from gamma ray data}}, \href{https://doi.org/10.1016/j.astropartphys.2011.06.010}{\emph{Astropart. Phys.} {\bfseries 35} (2011) 135} [\href{https://arxiv.org/abs/1012.5313}{{\ttfamily 1012.5313}}].

\bibitem{Taylor:2011bn}
A.M.~Taylor, I.~Vovk and A.~Neronov, \emph{{Extragalactic magnetic fields constraints from simultaneous GeV-TeV observations of blazars}}, \href{https://doi.org/10.1051/0004-6361/201116441}{\emph{Astron. Astrophys.} {\bfseries 529} (2011) A144} [\href{https://arxiv.org/abs/1101.0932}{{\ttfamily 1101.0932}}].

\bibitem{Takahashi:2013lba}
K.~Takahashi, M.~Mori, K.~Ichiki, S.~Inoue and H.~Takami, \emph{{Lower Bounds on Magnetic Fields in Intergalactic Voids from Long-term GeV-TeV Light Curves of the Blazar Mrk 421}}, \href{https://doi.org/10.1088/2041-8205/771/2/L42}{\emph{Astrophys. J. Lett.} {\bfseries 771} (2013) L42} [\href{https://arxiv.org/abs/1303.3069}{{\ttfamily 1303.3069}}].

\bibitem{Finke:2013tyq}
{\scshape Fermi-LAT} collaboration, \emph{{Constraints on the Intergalactic Magnetic Field from Gamma-Ray Observations of Blazars}}, {\emph{eConf} {\bfseries C121028} (2012) 365} [\href{https://arxiv.org/abs/1303.5093}{{\ttfamily 1303.5093}}].

\bibitem{Finke:2015ona}
J.D.~Finke, L.C.~Reyes, M.~Georganopoulos, K.~Reynolds, M.~Ajello, S.J.~Fegan et~al., \emph{{Constraints on the Intergalactic Magnetic Field with Gamma-Ray Observations of Blazars}}, \href{https://doi.org/10.1088/0004-637X/814/1/20}{\emph{Astrophys. J.} {\bfseries 814} (2015) 20} [\href{https://arxiv.org/abs/1510.02485}{{\ttfamily 1510.02485}}].

\bibitem{MAGIC:2022piy}
{\scshape MAGIC} collaboration, \emph{{A lower bound on intergalactic magnetic fields from time variability of 1ES 0229+200 from MAGIC and Fermi/LAT observations}}, \href{https://doi.org/10.1051/0004-6361/202244126}{\emph{Astron. Astrophys.} {\bfseries 670} (2023) A145} [\href{https://arxiv.org/abs/2210.03321}{{\ttfamily 2210.03321}}].

\bibitem{Turner:1987bw}
M.S.~Turner and L.M.~Widrow, \emph{{Inflation Produced, Large Scale Magnetic Fields}}, \href{https://doi.org/10.1103/PhysRevD.37.2743}{\emph{Phys. Rev. D} {\bfseries 37} (1988) 2743}.

\bibitem{Ratra:1991bn}
B.~Ratra, \emph{{Cosmological 'seed' magnetic field from inflation}}, \href{https://doi.org/10.1086/186384}{\emph{Astrophys. J. Lett.} {\bfseries 391} (1992) L1}.

\bibitem{Garretson:1992vt}
W.D.~Garretson, G.B.~Field and S.M.~Carroll, \emph{{Primordial magnetic fields from pseudoGoldstone bosons}}, \href{https://doi.org/10.1103/PhysRevD.46.5346}{\emph{Phys. Rev. D} {\bfseries 46} (1992) 5346} [\href{https://arxiv.org/abs/hep-ph/9209238}{{\ttfamily hep-ph/9209238}}].

\bibitem{Martin:2007ue}
J.~Martin and J.~Yokoyama, \emph{{Generation of Large-Scale Magnetic Fields in Single-Field Inflation}}, \href{https://doi.org/10.1088/1475-7516/2008/01/025}{\emph{JCAP} {\bfseries 01} (2008) 025} [\href{https://arxiv.org/abs/0711.4307}{{\ttfamily 0711.4307}}].

\bibitem{Demozzi:2009fu}
V.~Demozzi, V.~Mukhanov and H.~Rubinstein, \emph{{Magnetic fields from inflation?}}, \href{https://doi.org/10.1088/1475-7516/2009/08/025}{\emph{JCAP} {\bfseries 08} (2009) 025} [\href{https://arxiv.org/abs/0907.1030}{{\ttfamily 0907.1030}}].

\bibitem{Kanno:2009ei}
S.~Kanno, J.~Soda and M.-a.~Watanabe, \emph{{Cosmological Magnetic Fields from Inflation and Backreaction}}, \href{https://doi.org/10.1088/1475-7516/2009/12/009}{\emph{JCAP} {\bfseries 12} (2009) 009} [\href{https://arxiv.org/abs/0908.3509}{{\ttfamily 0908.3509}}].

\bibitem{Bamba:2006ga}
K.~Bamba and M.~Sasaki, \emph{{Large-scale magnetic fields in the inflationary universe}}, \href{https://doi.org/10.1088/1475-7516/2007/02/030}{\emph{JCAP} {\bfseries 02} (2007) 030} [\href{https://arxiv.org/abs/astro-ph/0611701}{{\ttfamily astro-ph/0611701}}].

\bibitem{Anber:2006xt}
M.M.~Anber and L.~Sorbo, \emph{{N-flationary magnetic fields}}, \href{https://doi.org/10.1088/1475-7516/2006/10/018}{\emph{JCAP} {\bfseries 10} (2006) 018} [\href{https://arxiv.org/abs/astro-ph/0606534}{{\ttfamily astro-ph/0606534}}].

\bibitem{Campanelli:2013mea}
L.~Campanelli, \emph{{Origin of Cosmic Magnetic Fields}}, \href{https://doi.org/10.1103/PhysRevLett.111.061301}{\emph{Phys. Rev. Lett.} {\bfseries 111} (2013) 061301} [\href{https://arxiv.org/abs/1304.6534}{{\ttfamily 1304.6534}}].

\bibitem{Ferreira:2013sqa}
R.J.Z.~Ferreira, R.K.~Jain and M.S.~Sloth, \emph{{Inflationary magnetogenesis without the strong coupling problem}}, \href{https://doi.org/10.1088/1475-7516/2013/10/004}{\emph{JCAP} {\bfseries 10} (2013) 004} [\href{https://arxiv.org/abs/1305.7151}{{\ttfamily 1305.7151}}].

\bibitem{Ferreira:2014hma}
R.J.Z.~Ferreira, R.K.~Jain and M.S.~Sloth, \emph{{Inflationary Magnetogenesis without the Strong Coupling Problem II: Constraints from CMB anisotropies and B-modes}}, \href{https://doi.org/10.1088/1475-7516/2014/06/053}{\emph{JCAP} {\bfseries 06} (2014) 053} [\href{https://arxiv.org/abs/1403.5516}{{\ttfamily 1403.5516}}].

\bibitem{Ng:2015ewp}
K.-W.~Ng, S.-L.~Cheng and W.~Lee, \emph{{Inflationary dilaton-axion magnetogenesis}}, \href{https://doi.org/10.6122/CJP.20150909}{\emph{Chin. J. Phys.} {\bfseries 53} (2015) 110105} [\href{https://arxiv.org/abs/1409.2656}{{\ttfamily 1409.2656}}].

\bibitem{Fujita:2015iga}
T.~Fujita, R.~Namba, Y.~Tada, N.~Takeda and H.~Tashiro, \emph{{Consistent generation of magnetic fields in axion inflation models}}, \href{https://doi.org/10.1088/1475-7516/2015/05/054}{\emph{JCAP} {\bfseries 05} (2015) 054} [\href{https://arxiv.org/abs/1503.05802}{{\ttfamily 1503.05802}}].

\bibitem{Adshead:2016iae}
P.~Adshead, J.T.~Giblin, T.R.~Scully and E.I.~Sfakianakis, \emph{{Magnetogenesis from axion inflation}}, \href{https://doi.org/10.1088/1475-7516/2016/10/039}{\emph{JCAP} {\bfseries 10} (2016) 039} [\href{https://arxiv.org/abs/1606.08474}{{\ttfamily 1606.08474}}].

\bibitem{Kushwaha:2020nfa}
A.~Kushwaha and S.~Shankaranarayanan, \emph{{Helical magnetic fields from Riemann coupling}}, \href{https://doi.org/10.1103/PhysRevD.102.103528}{\emph{Phys. Rev. D} {\bfseries 102} (2020) 103528} [\href{https://arxiv.org/abs/2008.10825}{{\ttfamily 2008.10825}}].

\bibitem{Tripathy:2022iev}
S.~Tripathy, D.~Chowdhury, H.V.~Ragavendra, R.K.~Jain and L.~Sriramkumar, \emph{{Circumventing the challenges in the choice of the nonconformal coupling function in inflationary magnetogenesis}}, \href{https://doi.org/10.1103/PhysRevD.107.043501}{\emph{Phys. Rev. D} {\bfseries 107} (2023) 043501} [\href{https://arxiv.org/abs/2211.05834}{{\ttfamily 2211.05834}}].

\bibitem{Vachaspati:1991nm}
T.~Vachaspati, \emph{{Magnetic fields from cosmological phase transitions}}, \href{https://doi.org/10.1016/0370-2693(91)90051-Q}{\emph{Phys. Lett. B} {\bfseries 265} (1991) 258}.

\bibitem{Sigl:1996dm}
G.~Sigl, A.V.~Olinto and K.~Jedamzik, \emph{{Primordial magnetic fields from cosmological first order phase transitions}}, \href{https://doi.org/10.1103/PhysRevD.55.4582}{\emph{Phys. Rev. D} {\bfseries 55} (1997) 4582} [\href{https://arxiv.org/abs/astro-ph/9610201}{{\ttfamily astro-ph/9610201}}].

\bibitem{Dolgov:2001nv}
A.D.~Dolgov and D.~Grasso, \emph{{Generation of cosmic magnetic fields and gravitational waves at neutrino decoupling}}, \href{https://doi.org/10.1103/PhysRevLett.88.011301}{\emph{Phys. Rev. Lett.} {\bfseries 88} (2002) 011301} [\href{https://arxiv.org/abs/astro-ph/0106154}{{\ttfamily astro-ph/0106154}}].

\bibitem{Stevens:2007ep}
T.~Stevens, M.B.~Johnson, L.S.~Kisslinger, E.M.~Henley, W.Y.P.~Hwang and M.~Burkardt, \emph{{Role of Charged Gauge Fields in Generating Magnetic Seed Fields in Bubble Collisions during the Cosmological Electroweak Phase Transition}}, \href{https://doi.org/10.1103/PhysRevD.77.023501}{\emph{Phys. Rev. D} {\bfseries 77} (2008) 023501} [\href{https://arxiv.org/abs/0707.1346}{{\ttfamily 0707.1346}}].

\bibitem{DiazGil:2007dy}
A.~Diaz-Gil, J.~Garcia-Bellido, M.~Garcia~Perez and A.~Gonzalez-Arroyo, \emph{{Magnetic field production during preheating at the electroweak scale}}, \href{https://doi.org/10.1103/PhysRevLett.100.241301}{\emph{Phys. Rev. Lett.} {\bfseries 100} (2008) 241301} [\href{https://arxiv.org/abs/0712.4263}{{\ttfamily 0712.4263}}].

\bibitem{Kahniashvili:2009qi}
T.~Kahniashvili, A.G.~Tevzadze and B.~Ratra, \emph{{Phase Transition Generated Cosmological Magnetic Field at Large Scales}}, \href{https://doi.org/10.1088/0004-637X/726/2/78}{\emph{Astrophys. J.} {\bfseries 726} (2011) 78} [\href{https://arxiv.org/abs/0907.0197}{{\ttfamily 0907.0197}}].

\bibitem{Henley:2010ba}
E.M.~Henley, M.B.~Johnson and L.S.~Kisslinger, \emph{{EWPT Nucleation With MSSM and Electromagnetic Field Creation}}, \href{https://doi.org/10.1103/PhysRevD.81.085035}{\emph{Phys. Rev. D} {\bfseries 81} (2010) 085035} [\href{https://arxiv.org/abs/1001.2783}{{\ttfamily 1001.2783}}].

\bibitem{Grasso:2000wj}
D.~Grasso and H.R.~Rubinstein, \emph{{Magnetic fields in the early universe}}, \href{https://doi.org/10.1016/S0370-1573(00)00110-1}{\emph{Phys. Rept.} {\bfseries 348} (2001) 163} [\href{https://arxiv.org/abs/astro-ph/0009061}{{\ttfamily astro-ph/0009061}}].

\bibitem{Cheng:1994yr}
B.-l.~Cheng and A.V.~Olinto, \emph{{Primordial magnetic fields generated in the quark - hadron transition}}, \href{https://doi.org/10.1103/PhysRevD.50.2421}{\emph{Phys. Rev. D} {\bfseries 50} (1994) 2421}.

\bibitem{Durrer:2013pga}
R.~Durrer and A.~Neronov, \emph{{Cosmological Magnetic Fields: Their Generation, Evolution and Observation}}, \href{https://doi.org/10.1007/s00159-013-0062-7}{\emph{Astron. Astrophys. Rev.} {\bfseries 21} (2013) 62} [\href{https://arxiv.org/abs/1303.7121}{{\ttfamily 1303.7121}}].

\bibitem{Subramanian:2015lua}
K.~Subramanian, \emph{{The origin, evolution and signatures of primordial magnetic fields}}, \href{https://doi.org/10.1088/0034-4885/79/7/076901}{\emph{Rept. Prog. Phys.} {\bfseries 79} (2016) 076901} [\href{https://arxiv.org/abs/1504.02311}{{\ttfamily 1504.02311}}].

\bibitem{Vachaspati:2020blt}
T.~Vachaspati, \emph{{Progress on cosmological magnetic fields}}, \href{https://doi.org/10.1088/1361-6633/ac03a9}{\emph{Rept. Prog. Phys.} {\bfseries 84} (2021) 074901} [\href{https://arxiv.org/abs/2010.10525}{{\ttfamily 2010.10525}}].

\bibitem{Nandi:2017ajk}
D.~Nandi and S.~Shankaranarayanan, \emph{{Vector Galileon and inflationary magnetogenesis}}, \href{https://doi.org/10.1088/1475-7516/2018/01/039}{\emph{JCAP} {\bfseries 01} (2018) 039} [\href{https://arxiv.org/abs/1704.06897}{{\ttfamily 1704.06897}}].

\bibitem{Sharma:2017eps}
R.~Sharma, S.~Jagannathan, T.R.~Seshadri and K.~Subramanian, \emph{{Challenges in Inflationary Magnetogenesis: Constraints from Strong Coupling, Backreaction and the Schwinger Effect}}, \href{https://doi.org/10.1103/PhysRevD.96.083511}{\emph{Phys. Rev. D} {\bfseries 96} (2017) 083511} [\href{https://arxiv.org/abs/1708.08119}{{\ttfamily 1708.08119}}].

\bibitem{Sharma:2018kgs}
R.~Sharma, K.~Subramanian and T.R.~Seshadri, \emph{{Generation of helical magnetic field in a viable scenario of inflationary magnetogenesis}}, \href{https://doi.org/10.1103/PhysRevD.97.083503}{\emph{Phys. Rev. D} {\bfseries 97} (2018) 083503} [\href{https://arxiv.org/abs/1802.04847}{{\ttfamily 1802.04847}}].

\bibitem{Brandenberger:2025gks}
R.~Brandenberger, J.~Fr{\"o}hlich and H.~Jiao, \emph{{Cosmological Magnetic Fields from Ultralight Dark Matter}}, \href{https://doi.org/10.1103/ys32-853g}{\emph{Phys. Rev. Lett.} {\bfseries 136} (2026) 031001} [\href{https://arxiv.org/abs/2502.19310}{{\ttfamily 2502.19310}}].

\bibitem{Brahma:2026fre}
N.~Brahma and R.~Brandenberger, \emph{{Parametric Resonance and Backreaction Effects in Magnetogenesis from Ultralight Dark Matter}},  \href{https://arxiv.org/abs/2602.04285}{{\ttfamily 2602.04285}}.

\bibitem{Kushwaha:2026aab}
A.~Kushwaha and R.~Brandenberger, \emph{{ALP dark matter, cosmological magnetic fields, and the direct collapse black hole formation scenario}}, \href{https://doi.org/10.1103/hp1n-2jy4}{\emph{Phys. Rev. D} {\bfseries 114} (2026) 083515} [\href{https://arxiv.org/abs/2602.22170}{{\ttfamily 2602.22170}}].

\bibitem{Sharma:2026hzp}
R.~Sharma, S.~Majumdar and D.~Sachdeva, \emph{{Suppressed Magnetogenesis from Ultralight Dark Matter due to Finite Conductivity}},  \href{https://arxiv.org/abs/2604.17230}{{\ttfamily 2604.17230}}.

\bibitem{Gerlach:2026sga}
C.~Gerlach, D.~Schmitt and P.~Schwaller, \emph{{Audible Axion Magnetogenesis: Linking Intergalactic Magnetic Fields and Gravitational Waves}},  \href{https://arxiv.org/abs/2605.21092}{{\ttfamily 2605.21092}}.

\bibitem{Adshead:2015pva}
P.~Adshead, J.T.~Giblin, T.R.~Scully and E.I.~Sfakianakis, \emph{{Gauge-preheating and the end of axion inflation}}, \href{https://doi.org/10.1088/1475-7516/2015/12/034}{\emph{JCAP} {\bfseries 12} (2015) 034} [\href{https://arxiv.org/abs/1502.06506}{{\ttfamily 1502.06506}}].

\bibitem{Kamarpour:2018ckk}
M.~Kamarpour and O.~Sobol, \emph{{Magnetogenesis in Natural Inflation Model}}, \href{https://doi.org/10.15407/ujpe63.8.673}{\emph{Ukr. J. Phys.} {\bfseries 63} (2018) 673}.

\bibitem{Sobol:2019xls}
O.O.~Sobol, E.V.~Gorbar and S.I.~Vilchinskii, \emph{{Backreaction of electromagnetic fields and the Schwinger effect in pseudoscalar inflation magnetogenesis}}, \href{https://doi.org/10.1103/PhysRevD.100.063523}{\emph{Phys. Rev. D} {\bfseries 100} (2019) 063523} [\href{https://arxiv.org/abs/1907.10443}{{\ttfamily 1907.10443}}].

\bibitem{Gorbar:2021zlr}
E.V.~Gorbar, K.~Schmitz, O.O.~Sobol and S.I.~Vilchinskii, \emph{{Hypermagnetogenesis from axion inflation: Model-independent estimates}}, \href{https://doi.org/10.1103/PhysRevD.105.043530}{\emph{Phys. Rev. D} {\bfseries 105} (2022) 043530} [\href{https://arxiv.org/abs/2111.04712}{{\ttfamily 2111.04712}}].

\bibitem{Gorbar:2021rlt}
E.V.~Gorbar, K.~Schmitz, O.O.~Sobol and S.I.~Vilchinskii, \emph{{Gauge-field production during axion inflation in the gradient expansion formalism}}, \href{https://doi.org/10.1103/PhysRevD.104.123504}{\emph{Phys. Rev. D} {\bfseries 104} (2021) 123504} [\href{https://arxiv.org/abs/2109.01651}{{\ttfamily 2109.01651}}].

\bibitem{Brandenburg:2024awd}
A.~Brandenburg, O.~Iarygina, E.I.~Sfakianakis and R.~Sharma, \emph{{Magnetogenesis from axion-SU(2) inflation}}, \href{https://doi.org/10.1088/1475-7516/2024/12/057}{\emph{JCAP} {\bfseries 12} (2024) 057} [\href{https://arxiv.org/abs/2408.17413}{{\ttfamily 2408.17413}}].

\bibitem{Sharma:2024nfu}
R.~Sharma, A.~Brandenburg, K.~Subramanian and A.~Vikman, \emph{{Lattice simulations of axion-U(1) inflation: gravitational waves, magnetic fields, and black holes}},  \href{https://arxiv.org/abs/2411.04854}{{\ttfamily 2411.04854}}.

\bibitem{Iarygina:2025ncl}
O.~Iarygina, E.I.~Sfakianakis and A.~Brandenburg, \emph{{Schwinger effect in axion inflation on a lattice}}, \href{https://doi.org/10.1103/1m23-ll9t}{\emph{Phys. Rev. D} {\bfseries 114} (2026) 063531} [\href{https://arxiv.org/abs/2506.20538}{{\ttfamily 2506.20538}}].

\bibitem{Kamada:2016cnb}
K.~Kamada and A.J.~Long, \emph{{Evolution of the Baryon Asymmetry through the Electroweak Crossover in the Presence of a Helical Magnetic Field}}, \href{https://doi.org/10.1103/PhysRevD.94.123509}{\emph{Phys. Rev. D} {\bfseries 94} (2016) 123509} [\href{https://arxiv.org/abs/1610.03074}{{\ttfamily 1610.03074}}].

\bibitem{Kamada:2020bmb}
K.~Kamada, F.~Uchida and J.~Yokoyama, \emph{{Baryon isocurvature constraints on the primordial hypermagnetic fields}}, \href{https://doi.org/10.1088/1475-7516/2021/04/034}{\emph{JCAP} {\bfseries 04} (2021) 034} [\href{https://arxiv.org/abs/2012.14435}{{\ttfamily 2012.14435}}].

\bibitem{Boyer:2025jno}
T.~Boyer and A.~Neronov, \emph{{Baryon asymmetry constraints on magnetic field from the Electroweak epoch}},  \href{https://arxiv.org/abs/2504.07937}{{\ttfamily 2504.07937}}.

\bibitem{Hamada:2025ooy}
Y.~Hamada, K.~Mukaida and F.~Uchida, \emph{{Revisiting constraints on magnetogenesis from baryon asymmetry}}, \href{https://doi.org/10.1103/45lx-vp2h}{\emph{Phys. Rev. D} {\bfseries 113} (2026) 083522} [\href{https://arxiv.org/abs/2509.23858}{{\ttfamily 2509.23858}}].

\bibitem{Yanagihara:2023qvx}
K.~Yanagihara, F.~Uchida, T.~Fujita and S.~Tsujikawa, \emph{{Low-scale inflationary magnetogenesis without baryon isocurvature problem}}, \href{https://doi.org/10.1103/PhysRevD.110.083506}{\emph{Phys. Rev. D} {\bfseries 110} (2024) 083506} [\href{https://arxiv.org/abs/2312.07938}{{\ttfamily 2312.07938}}].

\bibitem{Barnaby:2011vw}
N.~Barnaby, R.~Namba and M.~Peloso, \emph{{Phenomenology of a Pseudo-Scalar Inflaton: Naturally Large Nongaussianity}}, \href{https://doi.org/10.1088/1475-7516/2011/04/009}{\emph{JCAP} {\bfseries 04} (2011) 009} [\href{https://arxiv.org/abs/1102.4333}{{\ttfamily 1102.4333}}].

\bibitem{Allahverdi:2006iq}
R.~Allahverdi, K.~Enqvist, J.~Garcia-Bellido and A.~Mazumdar, \emph{{Gauge invariant MSSM inflaton}}, \href{https://doi.org/10.1103/PhysRevLett.97.191304}{\emph{Phys. Rev. Lett.} {\bfseries 97} (2006) 191304} [\href{https://arxiv.org/abs/hep-ph/0605035}{{\ttfamily hep-ph/0605035}}].

\bibitem{Allahverdi:2007wb}
R.~Allahverdi, K.~Enqvist, J.~Garcia-Bellido, A.~Jokinen and A.~Mazumdar, \emph{{MSSM flat direction inflation: Slow roll, stability, fine tuning and reheating}}, \href{https://doi.org/10.1088/1475-7516/2007/06/019}{\emph{JCAP} {\bfseries 06} (2007) 019} [\href{https://arxiv.org/abs/hep-ph/0610134}{{\ttfamily hep-ph/0610134}}].

\bibitem{Martin:2013nzq}
J.~Martin, C.~Ringeval and V.~Vennin, \emph{{Encyclopaedia Inflationaris}}, \href{https://doi.org/10.1016/j.dark.2014.01.003}{\emph{Phys. Dark Univ.} {\bfseries 5-6} (2014) 75} [\href{https://arxiv.org/abs/1303.3787}{{\ttfamily 1303.3787}}].

\bibitem{Barnaby:2010vf}
N.~Barnaby and M.~Peloso, \emph{{Large Nongaussianity in Axion Inflation}}, \href{https://doi.org/10.1103/PhysRevLett.106.181301}{\emph{Phys. Rev. Lett.} {\bfseries 106} (2011) 181301} [\href{https://arxiv.org/abs/1011.1500}{{\ttfamily 1011.1500}}].

\bibitem{Anber:2009ua}
M.M.~Anber and L.~Sorbo, \emph{{Naturally inflating on steep potentials through electromagnetic dissipation}}, \href{https://doi.org/10.1103/PhysRevD.81.043534}{\emph{Phys. Rev. D} {\bfseries 81} (2010) 043534} [\href{https://arxiv.org/abs/0908.4089}{{\ttfamily 0908.4089}}].

\bibitem{Caravano:2022epk}
A.~Caravano, E.~Komatsu, K.D.~Lozanov and J.~Weller, \emph{{Lattice simulations of axion-U(1) inflation}}, \href{https://doi.org/10.1103/PhysRevD.108.043504}{\emph{Phys. Rev. D} {\bfseries 108} (2023) 043504} [\href{https://arxiv.org/abs/2204.12874}{{\ttfamily 2204.12874}}].

\bibitem{Figueroa:2023oxc}
D.G.~Figueroa, J.~Lizarraga, A.~Urio and J.~Urrestilla, \emph{{Strong Backreaction Regime in Axion Inflation}}, \href{https://doi.org/10.1103/PhysRevLett.131.151003}{\emph{Phys. Rev. Lett.} {\bfseries 131} (2023) 151003} [\href{https://arxiv.org/abs/2303.17436}{{\ttfamily 2303.17436}}].

\bibitem{Figueroa:2024rkr}
D.G.~Figueroa, J.~Lizarraga, N.~Loayza, A.~Urio and J.~Urrestilla, \emph{{The non-linear dynamics of axion inflation: a detailed lattice study}},  \href{https://arxiv.org/abs/2411.16368}{{\ttfamily 2411.16368}}.

\bibitem{JOSS}
{Pencil Code Collaboration}, A.~{Brandenburg}, A.~{Johansen}, P.~{Bourdin}, W.~{Dobler}, W.~{Lyra} et~al., \emph{{The Pencil Code, a modular MPI code for partial differential equations and particles: multipurpose and multiuser-maintained}}, \href{https://doi.org/10.21105/joss.02807}{\emph{J. Open Source Software} {\bfseries 6} (2021) 2807}.

\bibitem{2006physics...5038H}
W.~{Heisenberg} and H.~{Euler}, \emph{{Consequences of Dirac Theory of the Positron}}, \href{https://doi.org/10.48550/arXiv.physics/0605038}{\emph{arXiv e-prints} (2006) physics/0605038} [\href{https://arxiv.org/abs/physics/0605038}{{\ttfamily physics/0605038}}].

\bibitem{Schwinger:1951nm}
J.S.~Schwinger, \emph{{On gauge invariance and vacuum polarization}}, \href{https://doi.org/10.1103/PhysRev.82.664}{\emph{Phys. Rev.} {\bfseries 82} (1951) 664}.

\bibitem{Frob:2014zka}
M.B.~Fr{\"o}b, J.~Garriga, S.~Kanno, M.~Sasaki, J.~Soda, T.~Tanaka et~al., \emph{{Schwinger effect in de Sitter space}}, \href{https://doi.org/10.1088/1475-7516/2014/04/009}{\emph{JCAP} {\bfseries 04} (2014) 009} [\href{https://arxiv.org/abs/1401.4137}{{\ttfamily 1401.4137}}].

\bibitem{Sharma:2017ivh}
R.~Sharma and S.~Singh, \emph{{Multifaceted Schwinger effect in de Sitter space}}, \href{https://doi.org/10.1103/PhysRevD.96.025012}{\emph{Phys. Rev. D} {\bfseries 96} (2017) 025012} [\href{https://arxiv.org/abs/1704.05076}{{\ttfamily 1704.05076}}].

\bibitem{Kobayashi:2014zza}
T.~Kobayashi and N.~Afshordi, \emph{{Schwinger Effect in 4D de Sitter Space and Constraints on Magnetogenesis in the Early Universe}}, \href{https://doi.org/10.1007/JHEP10(2014)166}{\emph{JHEP} {\bfseries 10} (2014) 166} [\href{https://arxiv.org/abs/1408.4141}{{\ttfamily 1408.4141}}].

\bibitem{Bavarsad:2016cxh}
E.~Bavarsad, C.~Stahl and S.-S.~Xue, \emph{{Scalar current of created pairs by Schwinger mechanism in de Sitter spacetime}}, \href{https://doi.org/10.1103/PhysRevD.94.104011}{\emph{Phys. Rev. D} {\bfseries 94} (2016) 104011} [\href{https://arxiv.org/abs/1602.06556}{{\ttfamily 1602.06556}}].

\bibitem{Domcke:2018eki}
V.~Domcke and K.~Mukaida, \emph{{Gauge Field and Fermion Production during Axion Inflation}}, \href{https://doi.org/10.1088/1475-7516/2018/11/020}{\emph{JCAP} {\bfseries 11} (2018) 020} [\href{https://arxiv.org/abs/1806.08769}{{\ttfamily 1806.08769}}].

\bibitem{vonEckardstein:2024tix}
R.~von Eckardstein, K.~Schmitz and O.~Sobol, \emph{{On the Schwinger effect during axion inflation}}, \href{https://doi.org/10.1007/JHEP02(2025)096}{\emph{JHEP} {\bfseries 02} (2025) 096} [\href{https://arxiv.org/abs/2408.16538}{{\ttfamily 2408.16538}}].

\bibitem{Banerjee:2004df}
R.~Banerjee and K.~Jedamzik, \emph{{The Evolution of cosmic magnetic fields: From the very early universe, to recombination, to the present}}, \href{https://doi.org/10.1103/PhysRevD.70.123003}{\emph{Phys. Rev. D} {\bfseries 70} (2004) 123003} [\href{https://arxiv.org/abs/astro-ph/0410032}{{\ttfamily astro-ph/0410032}}].

\bibitem{Brandenburg:2016odr}
A.~Brandenburg and T.~Kahniashvili, \emph{{Classes of hydrodynamic and magnetohydrodynamic turbulent decay}}, \href{https://doi.org/10.1103/PhysRevLett.118.055102}{\emph{Phys. Rev. Lett.} {\bfseries 118} (2017) 055102} [\href{https://arxiv.org/abs/1607.01360}{{\ttfamily 1607.01360}}].

\bibitem{Brandenburg:2017neh}
A.~Brandenburg, T.~Kahniashvili, S.~Mandal, A.~Roper~Pol, A.G.~Tevzadze and T.~Vachaspati, \emph{{Evolution of hydromagnetic turbulence from the electroweak phase transition}}, \href{https://doi.org/10.1103/PhysRevD.96.123528}{\emph{Phys. Rev. D} {\bfseries 96} (2017) 123528} [\href{https://arxiv.org/abs/1711.03804}{{\ttfamily 1711.03804}}].

\bibitem{Zhou:2019hkc}
M.~Zhou, P.~Bhat, N.F.~Loureiro and D.A.~Uzdensky, \emph{{Magnetic island merger as a mechanism for inverse magnetic energy transfer}}, \href{https://doi.org/10.1103/PhysRevResearch.1.012004}{\emph{Phys. Rev. Research.} {\bfseries 1} (2019) 012004} [\href{https://arxiv.org/abs/1901.02448}{{\ttfamily 1901.02448}}].

\bibitem{Bhat:2020anu}
P.~Bhat, M.~Zhou and N.F.~Loureiro, \emph{{Inverse energy transfer in decaying, three-dimensional, non-helical magnetic turbulence due to magnetic reconnection}}, \href{https://doi.org/10.1093/mnras/staa3849}{\emph{Mon. Not. Roy. Astron. Soc.} {\bfseries 501} (2021) 3074} [\href{https://arxiv.org/abs/2007.07325}{{\ttfamily 2007.07325}}].

\bibitem{Hosking:2020wom}
D.N.~Hosking and A.A.~Schekochihin, \emph{{Reconnection-Controlled Decay of Magnetohydrodynamic Turbulence and the Role of Invariants}}, \href{https://doi.org/10.1103/PhysRevX.11.041005}{\emph{Phys. Rev. X} {\bfseries 11} (2021) 041005} [\href{https://arxiv.org/abs/2012.01393}{{\ttfamily 2012.01393}}].

\bibitem{Zhou:2022xhk}
H.~Zhou, R.~Sharma and A.~Brandenburg, \emph{{Scaling of the Hosking integral in decaying magnetically dominated turbulence}}, \href{https://doi.org/10.1017/S002237782200109X}{\emph{J. Plasma Phys.} {\bfseries 88} (2022) 905880602} [\href{https://arxiv.org/abs/2206.07513}{{\ttfamily 2206.07513}}].

\bibitem{2024OJAp....7E..75D}
S.~{Dwivedi}, C.~{Anandavijayan} and P.~{Bhat}, \emph{{Quasi-two-dimensionality of three-dimensional, magnetically dominated, decaying turbulence}}, \href{https://doi.org/10.33232/001c.122855}{\emph{The Open Journal of Astrophysics} {\bfseries 7} (2024) 75} [\href{https://arxiv.org/abs/2401.01965}{{\ttfamily 2401.01965}}].

\bibitem{Kasselmann:2026tiu}
L.~Kasselmann, P.~Grete, P.~Trivedi, M.~Br{\"u}ggen and R.~Banerjee, \emph{{Inverse energy transfer in decaying MHD turbulence: A shell-to-shell analysis}},  \href{https://arxiv.org/abs/2606.02617}{{\ttfamily 2606.02617}}.

\bibitem{Hosking:2022umv}
D.N.~Hosking and A.A.~Schekochihin, \emph{{Cosmic-void observations reconciled with primordial magnetogenesis}}, \href{https://doi.org/10.1038/s41467-023-43258-3}{\emph{Nature Commun.} {\bfseries 14} (2023) 7523} [\href{https://arxiv.org/abs/2203.03573}{{\ttfamily 2203.03573}}].

\bibitem{Brandenburg:2024tyi}
A.~Brandenburg, A.~Neronov and F.~Vazza, \emph{{Resistively controlled primordial magnetic turbulence decay}}, \href{https://doi.org/10.1051/0004-6361/202449267}{\emph{Astron. Astrophys.} {\bfseries 687} (2024) A186} [\href{https://arxiv.org/abs/2401.08569}{{\ttfamily 2401.08569}}].

\bibitem{2026arXiv260117955A}
C.~{Anandavijayan} and P.~{Bhat}, \emph{{Spectral Evolution and Current Sheet Analysis as Probes of Reconnection-Mediated Decay in Magnetically Dominated Turbulence}}, \href{https://doi.org/10.48550/arXiv.2601.17955}{\emph{arXiv e-prints} (2026) arXiv:2601.17955} [\href{https://arxiv.org/abs/2601.17955}{{\ttfamily 2601.17955}}].

\end{thebibliography}\endgroup
\end{document}